\documentclass[11pt,a4paper]{article}
\usepackage[caption=false]{subfig}
\usepackage{hyperref}
\usepackage{multirow}
\usepackage[square,numbers,sort&compress]{natbib}
\usepackage{color}
\newcommand{\non}{\nonumber\\}
\usepackage{amsmath,amssymb}
\usepackage{graphicx}
\usepackage[normalem]{ulem} 
\usepackage{cancel} 

\renewcommand\sout{\bgroup \color{red} \ULdepth=-.5ex \ULset}

\newcommand{\del}{\partial}

\begin{document}

\begin{titlepage}
\null
\begin{flushright}
November, 2018
\end{flushright}
\vskip 2.5cm
\begin{center}
{\LARGE \bf 
Temporally, Spatially or Light-like Modulated Vacua 
\\
\vskip 0.3cm
in Lorentz Invariant Theories
}
\vskip 1.7cm
\normalsize
\renewcommand\thefootnote{\alph{footnote}}

{\large
Sven Bjarke Gudnason$^{\dagger}$\footnote{gudnason(at)keio.jp}, 
Muneto Nitta$^{\dagger}$\footnote{nitta(at)phys-h.keio.ac.jp},
Shin Sasaki$^\ddagger$\footnote{shin-s(at)kitasato-u.ac.jp}
and 
Ryo Yokokura$^\dagger$\footnote{ryokokur(at)keio.jp}
}
\vskip 0.7cm
{\it
$^\dagger$ 
Department of Physics, and Research and Education Center for Natural Sciences, \\
\vskip -0.2cm
Keio University, Hiyoshi 4-1-1, Yokohama, Kanagawa 223-8521, Japan  
\vskip 0.1cm
$^\ddagger$
Department of Physics,  Kitasato University, Sagamihara 252-0373, Japan
}
\vskip 0.5cm
\begin{abstract}
We study the vacuum structure of a class of Lorentz invariant field
theories where the vacuum expectation values are not constant but are
(phase) modulated. 
The vacua are classified into spatial, temporal, and
light-like modulation types according to the patterns of spontaneous
breaking of translational symmetry.
The conditions for having temporal or light-like modulated vacua  
imply severe constraints on the models.
We utilize the notion of generalized Nambu-Goldstone modes
which appear in the modulated vacua. 
Finally, we examine fluctuation modes around these vacua and discuss
their dynamics and the absence of ghosts.
\end{abstract}
\end{center}

\end{titlepage}

\newpage

\setcounter{footnote}{0}
\renewcommand\thefootnote{\arabic{footnote}}
\pagenumbering{arabic}

\tableofcontents


\section{Introduction} \label{sect:introduction}

Finding the vacuum structure is one of the most fundamental 
issues in the course of understanding quantum field theories.
The true (false) vacuum in field theories is defined by being 
the global (a local) minimum of the energy functional (Hamiltonian).
For models that consist of a canonical kinetic term (quadratic with 
second-order spacetime derivative) with a potential $V$, this is 
given by a constant field configuration determined by a minimum of
$V$. 
In this case, the constant vacuum expectation value (VEV) trivially
satisfies the equation of motion and the constant field configurations
automatically yield the minimum energy since varying a field costs
gradient energy.
Contrary to this, the situation can generally become more involved
when we consider models that contain quartic or higher-order kinetic
terms in addition to the canonical one, but only if not all terms
appear in the energy functional with a positive sign. 
Such models have terms with more than two spacetime derivatives and
are called higher-derivative models. 
In some cases, non-constant field configurations can be energetically
favored and become the vacuum.
Depending on the symmetries of the potential, the energy may be
minimized by a field configuration that has non-vanishing derivatives
in the vacuum (lowest-energy state).
The symmetry that will play an important role in this paper is shift
symmetry, which when possessed by the potential allows for the field
to be non-constant.
In this situation, the kinetic term and the higher-order derivative
terms can be seen as a ``potential'' for the ``velocity'' of the
field, that is, for a non-vanishing gradient of the field.
As mentioned above, if not all terms enter the energy functional with
a positive sign, the ``velocity'' that minimizes the energy may be a
non-vanishing constant.
Even if the canonical kinetic term has the usual positive sign in the
energy functional, a constant VEV may be just a local vacuum, and
instead a \emph{modulated vacuum} becomes the global vacuum.
In this paper we will study the situation where the ``velocity'' is
the phase of a complex field, although that is not necessarily the
only possibility in general.

In condensed matter physics as opposed to relativistic quantum field
theory, the lowest-energy state is not the vacuum but is called the
ground state. 
Modulated ground states were originally discovered in non-relativistic 
theories, {\it i.e.}~theories without Lorentz invariance, in the context
of superconductors.
In such theories, spatially-modulated ground states can easily be
found. 
For example, Fulde-Ferrell-Larkin-Ovchinnikov (FFLO) states in
superconductors were proposed a long time
ago \cite{Fulde:1964zz,larkin:1964zz}.  
For the Fulde-Ferrell (FF) state \cite{Fulde:1964zz}, the phase of a
condensate in the ground state is spatially modulated along one
spatial direction, while for the Larkin-Ovchinnikov
(LO) \cite{larkin:1964zz}, the amplitude of the condensate is
modulated. 
The LO state is the ground state when an external magnetic field is
applied \cite{Machida:1984zz} (see also Ref.~\cite{Yoshii:2011yt}), 
while the FF state can be
the ground state when a magnetic field penetrating a superconducting
ring is present (see e.g.~Ref.~\cite{Yoshii:2014fwa}).
Recent experiments reveal that these FFLO states are realized in
several physical
systems \cite{FFLO-review,Radzihovsky:2010,Radzihovsky:2010b,Liao:2010}. 
In addition to condensed matter physics, the FFLO states appear in
a variety of fields. 
In the context of QCD, spatially-modulated chiral condensates of the
FF-type, called a dual chiral density wave or chiral
spiral \cite{Nakano:2004cd,Karasawa:2013zsa}, and the LO-type,  
called a real kink crystal \cite{Nickel:2009ke,Nickel:2009wj}, 
were contemplated to appear in a certain region of the QCD phase
diagram. 
In the Gross-Neveu or Nambu-Jona Lasino model in 1+1 dimensions,  
FFLO states for which both the phase and the amplitude are modulated, 
called twisted kink crystals \cite{Basar:2008im,Basar:2008ki},
were discussed at finite temperature and density \cite{Basar:2009fg}. 
The related issues were also discussed in the context of the AdS/CFT
correspondence
\cite{Nakamura:2009tf,Amoretti:2016bxs,Amoretti:2017axe,Amoretti:2017frz,Andrade:2017leb,Cai:2017qdz}.

On the other hand, temporal modulations have also been studied recently.
The idea of the time-crystal \cite{Shapere:2012nq,Wilczek:2012jt}
proposes a time-dependent ground state in which the continuous
translational symmetry along the time direction is spontaneously
broken to a discrete subgroup. 
This is an analogue of the ordinary crystalline structure of matter in
spatial directions, namely, a conventional crystal.
Although the time-crystal cannot be realized in thermal equilibrium
states \cite{Watanabe:2014hea}, it has been reported to be
experimentally realized in a non-equilibrium
state \cite{Zhang:2016,Choi:2016}. 
A diffusive Nambu-Goldstone mode corresponding to spontaneously broken
time-translation in a time-crystal was discussed in
Ref.~\cite{Hayata:2018qgt}. 
A temporally modulated state was also studied in
QCD \cite{Hayata:2013sea} and finite-density systems
\cite{Nicolis:2013sga}.

Further examples of time-dependent vacuum states have also been studied
in various contexts.
A famous example of such kind of vacuum 
is ghost condensation \cite{ArkaniHamed:2003uy}.
This proposal causes a modification of the long-range gravitational
force by means of a Higgs mechanism giving mass to the graviton and 
is studied mostly in the context of cosmology.
In ghost condensation, the VEV depends linearly on time but is not
modulated.

In all the examples in the literature that we mentioned so far,
{\it i.e.}~in condensed matter physics or in QCD, Lorentz invariance is
either absent or explicitly broken by finite density, temperature,
and/or an external magnetic field
with the exception of ghost condensation \cite{ArkaniHamed:2003uy}.
In ghost condensation, Lorentz invariance is not broken but the
kinetic term of the scalar field has the ``wrong'' sign, thus yielding
a ghost. 

On the other hand, modulated {\it vacua} in Lorentz invariant setups
with a normal kinetic term had not been studied until recently
\cite{Nitta:2017mgk,Nitta:2017yuf}.  
In refs.~\cite{Nitta:2017mgk, Nitta:2017yuf}, we realized such a
possibility in a simple Lorentz invariant model with
higher-derivative terms, by studying the vacuum structure possessing a
non-constant VEV.\footnote{In this model, the trivial vacuum at
  the origin becomes metastable, in contrast to the situation in
  ghost condensation where it becomes unstable.}
The VEV that we analyzed is a function of a specific direction in 
three-dimensional space. 
Such kind of vacuum is called spatially modulated.

In this paper, we study modulated vacua in Lorentz invariant field
theories.
As a continuation of our previous studies on spatially-modulated vacua 
\cite{Nitta:2017mgk,Nitta:2017yuf}, we here focus on temporally and
light-like modulated vacua.
Nevertheless, it proves instructive to include the spatially-modulated
case here so as to put them all on equal footing.
A temporally modulated vacuum is characterized by the property that the
time derivative of fields develop a VEV.
If the solution we are looking for is timelike -- due to the Lorentz
invariance of the class of theories we are considering -- we can
always set the spatial derivatives to zero without loss
of generality simply by a change of Lorentz frame. 
Similarly, the light-like modulation is defined by a VEV for
the derivative of fields on the light-cone.
A nontrivial problem for time-dependent solutions is that the
extrema of the Hamiltonian do not necessarily coincide with all
solutions to the Lagrangian equation of motion.
A subset of the solutions to the Lagrangian equations of motion, as we
will show in this paper, do yield extrema of the corresponding
Hamiltonian; these are the states we are looking for.
In the class of theories we consider in this paper, we will
furthermore assume that there is no dependence on the field,
$\varphi$, itself in the Lagrangian and that the Lagrangian only
depends on the first derivatives of the field, $\del_m \varphi$.

In restricting to a Lagrangian that only depends (nonlinearly) on the
first time derivative of the field, we sidestep the problem of the
Ostrogradsky instability and related ghost \cite{Ostrogradsky,Woodard:2006nt}. 
As in the ordinary vacuum in field theories, once the fields develop
nontrivial VEVs, some of the symmetries of the Lagrangian are
spontaneously broken. In the modulated states in non-relativistic
theories, there appears Nambu-Goldstone (NG) modes associated with
broken symmetries \cite{Lee:2015bva,Hidaka:2015xza,Hayata:2018qgt}. 
We will show that there is an analogue of NG modes (generalized NG
modes) in our models. The generalized NG modes correspond to the flat
direction of the ``potential'' whose quadratic kinetic term disappears 
in the Lagrangian for the case of spatial
modulation \cite{Nitta:2017mgk}. 
However, this turns out 
not to be the case for temporal or light-like modulation.

In this paper, we study the FF-type (phase) modulation
$\varphi\sim\exp(i\omega t)$. 
One may wonder that this modulation is rather usual in
non-relativistic theories. 
For instance, (nonlinear) Schr{\"o}dinger systems, 
$i\partial_t\varphi=\nabla\varphi+V\varphi $
$(i\partial_t\varphi =\nabla\varphi+g|\varphi|^2\varphi)$, 
have states with a time dependence of the above type. 
Therefore, such a time dependence is not usually called a temporal
modulation.  
Note, however, that this is not the case for Lorentz invariant
(relativistic) theories discussed in this paper.

The organization of this paper is as follows.
In the next section, we discuss the general conditions for modulated vacua.
We classify the vacua into the spatial, temporal and light-like types.
In section \ref{sect:NG_theorem}, we introduce the notion of generalized
NG modes and discuss the relation between them and the 
zero modes of the generalized mass matrix.
In section \ref{sect:6th_model}, by focusing on a higher-derivative
model, we show an example of the modulated vacua and the generalized NG
modes.
We also examine the dynamics of the fluctuation modes in the vacua and
discuss the absence of ghosts.
Section \ref{sect:conclusion} is devoted to a summary and discussion.

\section{General discussion on modulated vacua}
\label{sect:general_mod}

Throughout this paper, we consider Lorentz-invariant theories of 
a single complex scalar field $\varphi$.
We require that the theory does not suffer from the Ostrogradsky
instability \cite{Ostrogradsky,Woodard:2006nt}
 which can be caused by fourth or
higher order of spacetime derivative terms.
For simplicity of constructing models,
we assume that theories have shift symmetry of the fields and 
a global $U(1)$ symmetry:
\begin{align}
\varphi \to \varphi + c, \qquad 
\varphi \to e^{i \theta} \varphi,
\label{eq:symmetries}
\end{align}
with constant parameters $c$ and $\theta$.
The absence of the Ostrogradsky instability then implies that the
Lagrangian consists of $\partial_m \varphi, \partial_m\bar{\varphi}$
only: 
\begin{align}
\mathcal{L} = \mathcal{L} (\partial_m \varphi, \partial_m
 \bar{\varphi}).
\label{eq:Lagrangian}
\end{align}

In general, vacua are defined as field configurations that minimize the energy
density and they satisfy the equation of motion.
The conventional vacua, which are characterized by constant VEVs,
trivially satisfy these conditions.
However, if we look for vacua where VEVs are not constant but generally
spacetime dependent functions, the conditions are not trivially
satisfied.
In the following, we write down the general conditions for vacua in
this theory and discuss modulated vacua.

\subsection{Conditions for modulated vacua}

Starting from the Lagrangian \eqref{eq:Lagrangian}, the equation of
motion is 
\begin{align}
0 =& \ \del_m  \frac{\del \mathcal{L}}{\del (\del_m \varphi)} =
 \frac{\del^2 \mathcal{L}}{\del (\del_m \varphi) \del (\del_n \varphi)}
 \del_m \del_n \varphi + \frac{\del^2 \mathcal{L}}{\del (\del_m \varphi)
 \del (\del_n \bar{\varphi})} \del_m \del_n \bar{\varphi}, 
\notag \\
0 =& \ \del_m \frac{\del \mathcal{L}}{\del (\del_m \bar{\varphi})} =
 \frac{\del^2 \mathcal{L}}{\del (\del_m \bar{\varphi}) \del (\del_n
 \varphi)} \del_m \del_n \varphi + \frac{\del^2 \mathcal{L}}{\del
 (\del_m \bar{\varphi}) \del (\del_n \bar{\varphi})} \del_m \del_n
 \bar{\varphi}.
\end{align}
This is expressed in compact form as:
\begin{align}
& 
\mathbf{L}^{mn} 
\left(
\begin{array}{c}
\del_m \del_n \varphi \\
\del_m \del_n \bar{\varphi}
\end{array}
\right) = 
\mathbf{L}^{00} 
\left(
\begin{array}{c}
\ddot{\varphi} \\
\ddot{\bar{\varphi}}
\end{array}
\right)
+
\left(
\mathbf{L}^{0i} 
+
\mathbf{L}^{i0}  
\right)
\left(
\begin{array}{c}
\del_i \dot{\varphi} \\
\del_i \dot{\bar{\varphi}}
\end{array}
\right)
+
\mathbf{L}^{ij}
\left(
\begin{array}{c}
\del_i \del_j \varphi \\
\del_i \del_j \bar{\varphi}
\end{array}
\right)
=
0.
\label{eq:eom}
\end{align}
Here $\dot{\varphi} = \del_0 \varphi$ and so on and 
the $2 \times 2$ matrices $\mathbf{L}^{mn} \ (m,n=0,1,2,3)$ are
defined as  
\begin{align}
\mathbf{L}^{mn} 
\equiv  
\left(
\begin{array}{cc}
\frac{\del^2 \mathcal{L}}{\del (\del_m \bar{\varphi}) \del (\del_n
 \varphi)} & \frac{\del^2 \mathcal{L}}{\del (\del_m \bar{\varphi})
 \del (\del_n \bar{\varphi})} 
 \\
\frac{\del^2 \mathcal{L}}{\del (\del_m \varphi) \del (\del_n \varphi)} & 
\frac{\del^2 \mathcal{L}}{\del (\del_m \varphi) \del (\del_n \bar{\varphi})}
\end{array}
\right).
\label{eq:Lagrangian_matrices}
\end{align}
Note that only the symmetric part $\mathbf{L}^{(mn)}$ contributes to the
equation of motion and $\mathbf{L}^{\dagger 00} = \mathbf{L}^{00}$,
$\mathbf{L}^{\dagger 0i} = \mathbf{L}^{i0}, \mathbf{L}^{\dagger ij} =
\mathbf{L}^{ji}$.

Any global or meta-stable vacuum of the theory satisfies the equation of
motion and it should be a local minimum of the energy functional. 
The Hamiltonian (energy) density is defined by
\begin{align}
\mathcal{H} = \frac{\del \mathcal{L}}{\del \dot{\varphi}} \dot{\varphi}
 + \frac{\del \mathcal{L}}{\del \dot{\bar{\varphi}}} \dot{\bar{\varphi}}
 - \mathcal{L}.
\end{align}
The Hamiltonian depends only on the spacetime derivative of fields.
Therefore it can be considered as a potential for $\del_m \varphi$.
The extremal condition of the energy 
with respect to $\del_m \varphi$ is then
\begin{align}
\frac{\del \mathcal{H}}{\del (\del_m \varphi)} =& \ 
\frac{\del^2 \mathcal{L}}{\del \dot{\varphi} \del (\del_m \varphi)}
 \dot{\varphi}
+ \frac{\del \mathcal{L}}{\del \dot{\varphi}} \delta^m {}_0 +
 \frac{\del^2 \mathcal{L}}{\del \dot{\bar{\varphi}} \del (\del_m
 \varphi)} \dot{\bar{\varphi}} - \frac{\del \mathcal{L}}{\del (\del_m
 \varphi)} = 0,
\end{align}
and its complex conjugate.
More explicitly, we have 
\begin{align}
\frac{\del \mathcal{H}}{\del \dot{\varphi}} =& \ \frac{\del^2
 \mathcal{L}}{\del \dot{\varphi}^2} \dot{\varphi} + \frac{\del^2
 \mathcal{L}}{\del \dot{\bar{\varphi}} \del \dot{\varphi}}
 \dot{\bar{\varphi}} = 0, 
 \notag \\
\frac{\del \mathcal{H}}{\del (\del_i \varphi)} =& \ 
\frac{\del^2 \mathcal{L}}{\del \dot{\varphi} \del (\del_i \varphi)}
 \dot{\varphi} + \frac{\del^2 \mathcal{L}}{\del \dot{\bar{\varphi}} \del
 (\del_i \varphi)} \dot{\bar{\varphi}} - \frac{\del \mathcal{L}}{\del
 (\del_i \varphi)} = 0.
\label{eq:Hamiltonian_extremum}
\end{align}
These together with their complex conjugates are expressed as 
\begin{align}
\mathbf{L}^{00} 
\left(
\begin{array}{c}
\dot{\varphi} \\
\dot{\bar{\varphi}}
\end{array}
\right) = 0,
\qquad 
\mathbf{L}^{i0} 
\left(
\begin{array}{c}
\dot{\varphi} \\
\dot{\bar{\varphi}}
\end{array}
\right)
- 
\left(
\begin{array}{c}
\frac{\del \mathcal{L}}{\del (\del_i \bar{\varphi})}
 \\
\frac{\del \mathcal{L}}{\del (\del_i \varphi)}
\end{array}
\right) = 0.
\label{eq:energy_extremum}
\end{align}
One finds that the conventional vacuum configuration which is
characterized by a constant  
VEV trivially satisfies the equation of motion \eqref{eq:eom}
and the extremal condition of the energy density \eqref{eq:energy_extremum}.
However as we will see in section \ref{sect:6th_model},
there can be nontrivial solutions to these equations
\eqref{eq:eom}, \eqref{eq:energy_extremum} where parts of $\del_m \varphi$
develop nonzero VEV $\langle \del_m \varphi \rangle \not= 0$.

In order to examine whether a configuration that satisfies
eqs.~\eqref{eq:eom} and \eqref{eq:energy_extremum} is a stable vacuum, we
consider the fluctuation $\phi$ in the configuration.
We consider a shift in the field
$\varphi \to \langle \varphi \rangle + \phi$, or
equivalently $\del_{m} \varphi \to  \langle \del_{m}
 \varphi \rangle + \del_m \phi$.
After the shift in the fields, the Hamiltonian is expanded as 
\begin{align}
&
\mathcal{H} \left( \langle \del_m \varphi \rangle + \del_m \phi,
 \langle \del_m \bar{\varphi} \rangle + \del_m
 \bar{\phi} \right)
\notag \\
& = 
\mathcal{H}_0 + 
\left. \frac{\del \mathcal{H}}{\del (\del_m \varphi)} \right|_0 \del_m
 \phi
+ \left. \frac{\del \mathcal{H}}{\del (\del_m \bar{\varphi})} \right|_0
 \del_m \bar{\phi} + 
\frac{1}{2}
\vec{\phi}_m^{\dagger}
 \mathbf{M}^{mn}|_0 \vec{\phi}_n + \cdots.,
\label{eq:Hamiltonian_expansion}
\end{align}
where the symbol $|_0$ means that they are evaluated in the vacuum and 
we have defined the $2 \times 2$ matrices $\mathbf{M}^{mn}$ and the
vector $\vec{\phi}_m$ as:
\begin{align}
\mathbf{M}^{mn} \equiv 
\left(
\begin{array}{cc}
\frac{\del^2 \mathcal{H}}{\del (\del_m \bar{\varphi}) \del (\del_n
 \varphi)} & \frac{\del^2 \mathcal{H}}{\del (\del_m \bar{\varphi})
 \del (\del_n \bar{\varphi})} 
 \\
\frac{\del^2 \mathcal{H}}{\del (\del_m \varphi) \del (\del_n \varphi)} & 
\frac{\del^2 \mathcal{H}}{\del (\del_m \varphi) \del (\del_n \bar{\varphi})}
\end{array}
\right),
\qquad 
\vec{\phi}_m \equiv 
\left(
\begin{array}{c}
\del_m \phi  \\
\del_m \bar{\phi}
\end{array}
\right).
\end{align}
Note that $\mathbf{M}^{\dagger 00} = \mathbf{M}^{00}$, $\mathbf{M}^{\dagger
0i} = \mathbf{M}^{i0}$, $\mathbf{M}^{\dagger ij} = \mathbf{M}^{ji}$.
The second and the third terms in eq.~\eqref{eq:Hamiltonian_expansion}
vanish $\left. \frac{\del \mathcal{H}}{\del (\del_m \varphi)}
\right|_0 = \left. \frac{\del \mathcal{H}}{\del (\del_m\bar\varphi)}
\right|_0 = 0$ due to the extremal condition of the energy density \eqref{eq:Hamiltonian_extremum}.
The matrices $\mathbf{M}^{mn}$ are expressed by the Lagrangian matrices \eqref{eq:Lagrangian_matrices}:
\begin{align}
\mathbf{M}^{mn} = 
\frac{\del \mathbf{L}^{mn}}{\del \dot{\varphi}}
 \dot{\varphi}
+ \frac{\del \mathbf{L}^{mn}}{\del \dot{\bar{\varphi}}}
 \dot{\bar{\varphi}}
+ \mathbf{L}^{0n} \delta^m {}_0 + \mathbf{L}^{m0} \delta^n {}_0 -
 \mathbf{L}^{mn}.
\end{align}
More explicitly, we have
\begin{align}
\mathbf{M}^{00} =& \ 
\frac{\del \mathbf{L}^{00}}{\del \dot{\varphi}}
 \dot{\varphi}
+ \frac{\del \mathbf{L}^{00}}{\del \dot{\bar{\varphi}}}
 \dot{\bar{\varphi}}
+
 \mathbf{L}^{00}, 
\notag \\
\mathbf{M}^{0i} =& \ 
\frac{\del \mathbf{L}^{0i}}{\del \dot{\varphi}}
 \dot{\varphi}
+ \frac{\del \mathbf{L}^{0i}}{\del \dot{\bar{\varphi}}}
 \dot{\bar{\varphi}},
\notag \\
\mathbf{M}^{ij} =& \ 
\frac{\del \mathbf{L}^{ij}}{\del \dot{\varphi}}
 \dot{\varphi}
+ \frac{\del \mathbf{L}^{ij}}{\del \dot{\bar{\varphi}}}
 \dot{\bar{\varphi}}
 -  \mathbf{L}^{ij}.
\label{eq:M_components}
\end{align}

In order to define the Hessian matrix of the energy functional, we
rearrange the expression \eqref{eq:Hamiltonian_expansion}:
\begin{align}
\mathcal{H} = \mathcal{H}_0 + \frac{1}{2} \vec{\phi}^{\dagger}_m
 \mathbf{M}^{mn}|_0 \vec{\phi}_n = \mathcal{H}_0 + \frac{1}{2}
 \vec{\Phi}^{\dagger} \mathcal{M}|_0 \vec{\Phi},
\end{align}
where we have defined the vector 
\begin{align}
\vec{\Phi} \equiv 
\left(
\begin{array}{c}
\partial_0 \phi\\
\partial_0 \bar{\phi}\\
\partial_1 \phi\\
\vdots\\
\partial_3 \bar{\phi}
\end{array}
\right)
= \left(
\begin{array}{c}
\vec{\phi}_0\\
\vec{\phi}_1\\
\vdots\\
\vec{\phi}_3
\end{array}
\right),
\label{eq:generalized_vector}
\end{align}
and the generalized $2d \times 2d$ mass matrix
\begin{align}
\mathcal{M} \equiv 
\left(
\begin{array}{cccc}
\mathbf{M}^{00} & \mathbf{M}^{01} & \cdots & \mathbf{M}^{03}
 \\
\mathbf{M}^{10} & \mathbf{M}^{11} & \cdots & 
\\
\vdots & & \ddots & 
\\
\mathbf{M}^{30} & \cdots & & \mathbf{M}^{33} 
\end{array}
\right).
\label{eq:generalized_mass_matrix}
\end{align}
Since $\mathbf{M}^{\dagger mn} = \mathbf{M}^{nm}$, 
the $2d \times 2d = 8 \times 8$ matrix $\mathcal{M}$ is a Hermitian
Hessian matrix.
In order for a vacuum $\langle \del_m \varphi \rangle \not= 0$ to be
stable, the generalized mass matrix needs to be positive semi-definite in the
vacuum which we denote $\mathcal{M}|_0 \ge 0$.

Once a vacuum $\langle \partial_{\hat{m}} \varphi \rangle \not= 0$, for
fixed $\hat{m}$, that satisfies
eqs.~\eqref{eq:eom}, \eqref{eq:Hamiltonian_extremum} and $\mathcal{M}|_0 \ge
0$ is found, we consider the dynamical (fluctuation) field $\phi$ around the vacuum.
Then the Lagrangian for the dynamical field is
\begin{align}
\mathcal{L} =& \ \mathcal{L}_0 + 
\left. \frac{\del \mathcal{L}}{\del (\del \dot{\varphi})} \right|_0 \del_0
 \phi
+ \left. \frac{\del \mathcal{L}}{\del (\del \dot{\bar{\varphi}})} \right|_0
 \del_0 \bar{\phi}
+ \left. \frac{\del \mathcal{L}}{\del (\del_i \varphi)} \right|_0 \del_i
 \phi 
+ \left. \frac{\del \mathcal{L}}{\del (\del_i \bar{\varphi})} \right|_0
 \del_i \bar{\phi}
\notag \\
& 
+ \frac{1}{2} \vec{\phi}_0^{\dagger} \mathbf{L}^{00}|_0 \vec{\phi}_0
+ \frac{1}{2} \vec{\phi}_0^{\dagger} \mathbf{L}^{0i}|_0 \vec{\phi}_i
+ \frac{1}{2} \vec{\phi}_i^{\dagger} \mathbf{L}^{i0}|_0 \vec{\phi}_0
+ \frac{1}{2} \vec{\phi}_i^{\dagger} \mathbf{L}^{ij}|_0 \vec{\phi}_j
+ \cdots .
\label{eq:fluctuation_Lagrangian}
\end{align}
Here $\mathbf{L}^{mn}$ are the matrices defined in
eq.~\eqref{eq:Lagrangian_matrices} and $\mathcal{L}_0$ is a constant.

In the next subsection, we examine nontrivial solutions to the
conditions \eqref{eq:eom} and \eqref{eq:energy_extremum}.
Among other things, we focus on vacuum configurations where the VEVs
are modulated with respect to spacetime coordinates.
We first classify the modulated vacua and then clarify the conditions
for the vacua.

\subsection{Classification of modulated vacua}
\label{subsect:classification_mod_vac}

The modulated vacua are classified according to the patterns of  
spontaneous breaking of the global and Poincar\'e symmetries.
When some components of $\del_{\hat{m}} \varphi$ develop a nonzero VEV, 
they break translational symmetry 
\footnote{
Recently, the spontaneous breaking of the
translational symmetry in a higher derivative model without the
Ostrogradsky instability is discussed \cite{Musso:2018wbv}.
}
along the $x^{\hat{m}}$-directions
as well as the rotational symmetry in a plane spanned by the
$x^{\hat{m}}$ coordinate. 
The shift and global $U(1)$ symmetries are also spontaneously broken.
If the VEV is a Fulde-Ferrell (FF, the phase modulated) type,
{\it i.e.}
$\langle\del_{\hat{m}} \varphi \rangle \propto e^{i c x^{\hat{m}}}$
with a constant $c$, then the symmetry breaking pattern is classified
by the vector generators $P^{\hat{m}}$ associated with the
translational symmetry along the $x^{\hat{m}}$-direction into three
cases (1) space-like, (2) time-like and (3) light-like. 

\begin{itemize}
\item[(1)] When $P^{\hat{m}}$ is space-like, the vacuum is called spatially modulated.
In this case, we can choose $\hat{m}=1$ without loss of generality.
The symmetry is spontaneously broken as $ISO(1,3) \times U(1) \times S
\to ISO(1,2)\rtimes [U(1) \times \mathcal{P}^{1}]_{\text{diag}}$.
Here $S$ is the shift symmetry group, $\mathcal{P}^1$ is the
translational group generated by $P^1$ and diag.~stands for the diagonal
subgroup corresponding to the simultaneous transformation:
\begin{align}
x^1 \to x^1 + a, \qquad \varphi \to e^{-ica} \varphi,
\end{align}
and $\rtimes$ implies a semi-direct product.

\item[(2)] When $P^{\hat{m}}$ is time-like, the vacuum is called temporally
modulated. In this case, we can choose $\hat{m} = 0$ without loss of
generality. 
In this vacuum, the symmetry is spontaneously broken as
$ISO(1,3) \times U(1) \times S \to ISO(3) \rtimes [U(1) \times
\mathcal{P}^{0}]_{\text{diag}}$ where $ISO(3)$ is the Poincar\'e group in the
spatial directions
\footnote{
This kind of symmetry breaking in Lorentz-invariant theories has been discussed in Ref.~\cite{Nicolis:2011pv}.}.

\item[(3)] On the other hand, when $P^{\hat{m}}$ is light-like (null), 
we can choose $P^{\hat{m}}$ to be the light-cone directions $P^{\pm} =
P^{0} \pm P^1$.
We call the vacuum light-like modulated.
The breaking pattern is 
$ ISO(1,3) \times U(1) \times S \to ISO(2) \times [U(1) \times
\mathcal{P}^{\pm}]_{\text{diag}}$.
\end{itemize}

We will discuss a generalization of the NG modes associated with these symmetry
breakings in section \ref{sect:NG_theorem}.
In the following, we examine the conditions for each modulated vacuum.

\paragraph{Spatial modulation}
In the case of the spatial modulation, the VEV is characterized by
\begin{align}
\langle \dot{\varphi} \rangle = 
\langle \del_2 \varphi \rangle =
\langle \del_3 \varphi \rangle = 0, \qquad \langle \del_{1} \varphi \rangle
 \not= 0.
\label{eq:sp}
\end{align}
Assuming the Ansatz \eqref{eq:sp}, we examine the conditions
\eqref{eq:eom} and \eqref{eq:energy_extremum}.
By this assumption, we look for static field configurations.
We first demand the following conditions
\begin{align}
\left. \frac{\del \mathcal{L}}{\del (\del_i \varphi)}\right|_0 = 0,
 \quad (i=1,2,3).
\label{eq:sp1}
\end{align}
Note that $|_0$ means that the conditions are satisfied in the vacuum for any
$x^i$.
If we impose the above conditions, the energy extremal conditions in 
eq.~\eqref{eq:energy_extremum} are trivially satisfied.
The equation of motion \eqref{eq:eom}
is also trivially satisfied since $\varphi$ is $x^0$-independent and $\frac{\del \mathcal{L}}{\del (\del_i
\varphi)} = 0$ for any $x^i$.
Indeed, for a static field configuration,  we have the relation
$\mathcal{H} = - \mathcal{L}$ and the condition $\delta \mathcal{H} = 0$
automatically implies $\delta \mathcal{L} = 0$.
In refs.~\cite{Nitta:2017mgk, Nitta:2017yuf} we
solved the condition \eqref{eq:sp1} and found a spatially modulated
vacuum.

If the theory that we consider is Lorentz invariant, one notices that
only the combinations 
$\dot{\varphi}^2, \dot{\varphi} \dot{\bar{\varphi}}$ appear in the
Lagrangian:
\begin{align}
\mathcal{L} \supset A \dot{\varphi}^2, B \dot{\varphi} \dot{\bar{\varphi}},
\end{align}
where $A,B$ are any Lorentz invariant terms. Then, one can show that the relations, 
\begin{align}
\left. \frac{\del \mathcal{L}}{\del \dot{\varphi}}
 \right|_{\dot{\varphi} = 0} 
=
\left. 
\frac{\del^2 \mathcal{L}}{\del \dot{\varphi} \del (\del_i \varphi)}
\right|_{\dot{\varphi} = 0}
= 0,
\label{eq:sm_relation}
\end{align}
always hold for the spatial modulation Ansatz \eqref{eq:sp}.
From eq.~\eqref{eq:M_components}, we therefore find the relations
\begin{align}
\mathbf{M}^{00}|_0 = \mathbf{L}^{00}|_0, \quad 
\mathbf{M}^{ij}|_0 = - \mathbf{L}^{ij}|_0.
\label{eq:sp_Lagrangian_matrices}
\end{align}
In particular one can show that $\mathbf{L}^{0i}|_0 = \mathbf{L}^{i0}|_0 =
\mathbf{M}^{0i}|_0 = 0$ for the spatial modulation.
The generalized mass matrix is given by 
\begin{align}
\mathcal{M}|_0 = 
\left(
\begin{array}{c|c}
\mathbf{L}^{00}|_0 & 0 \\
\hline
0 & - \text{diag} (\mathbf{L}^{ij})|_0
\end{array}
\right).
\end{align}
This is completely block diagonal. 
The explicit expression for $\mathcal{M}$ depends on the models under
consideration. 
For the vacuum \eqref{eq:sp}, we require that $\mathcal{M}|_0$ is positive semi-definite.

Once a vacuum is found, the Lagrangian for the fluctuation in a
spatially modulated vacuum becomes
\begin{align}
\mathcal{L} = \mathcal{L}_0 + \frac{1}{2} \vec{\phi}_0^{\dagger} \mathbf{L}^{00}|_0
 \vec{\phi}_0
+ \frac{1}{2} \vec{\phi}_i^{\dagger} \mathbf{L}^{ij}|_0 \vec{\phi}_j + \cdots.
\end{align}
The matrices $\mathbf{L}^{mn}|_0$ govern the kinetic terms of the
fluctuation modes. For the spatial modulation, the zero modes of
$\mathbf{M}^{mn}|_0$ coincide with those of $\mathbf{L}^{mn}|_0$.
This issue will be discussed in section \ref{sect:NG_theorem}.

\paragraph{Temporal modulation}
The temporal modulation is characterized by the VEV:
\begin{align}
\langle 
\dot{\varphi}
\rangle \not= 0, \quad 
\langle \del_i \varphi \rangle = 0, \quad (i=1,2,3).
\label{eq:tm}
\end{align}
In the case of temporal modulations, it is not always true that extrema
of the energy are solutions to the equation of motion.
In order to look for a configuration characterized by
eq.~\eqref{eq:tm}, we demand the following conditions:
\begin{align}
\mathbf{L}^{00}|_0 = 0, \quad 
\left. \frac{\del \mathcal{L}}{\del (\del_i \varphi)} \right|_0 = 0.
\label{eq:temp_conditions}
\end{align}
Indeed, the second condition in eq.~\eqref{eq:temp_conditions} allows a solution 
\begin{align}
\del_i \varphi = 0,
\label{eq:temp_mod_cond}
\end{align}
in Lorentz invariant models. 
This substantially implies $\mathbf{L}^{0i}|_0 = 0$.
Since the solution \eqref{eq:temp_mod_cond} implies that $\partial_i \partial_j
\varphi = 0$ at any points, the conditions \eqref{eq:eom} and
\eqref{eq:energy_extremum} necessary for modulated vacua are now
satisfied.
By the Ansatz \eqref{eq:tm}, we have  
\begin{align}
\mathbf{M}^{00}|_0 =& \ 
\left(
\frac{\del \mathbf{L}^{00}}{\del \dot{\varphi}}
\dot{\varphi}
\right)_0 + 
\left(
\frac{\del \mathbf{L}^{00}}{\del \dot{\bar{\varphi}}}
\dot{\bar{\varphi}}
\right)_0, 
\notag \\
\mathbf{M}^{0i}|_0 =& \ 
\left(
\frac{\del \mathbf{L}^{0i}}{\del \dot{\varphi}}
 \dot{\varphi}
\right)_0
+ 
\left(
\frac{\del \mathbf{L}^{0i}}{\del \dot{\bar{\varphi}}}
 \dot{\bar{\varphi}}
\right)_0
,
\notag \\
\mathbf{M}^{ij}|_0 =& \ 
\left(
\frac{\del \mathbf{L}^{ij}}{\del \dot{\varphi}}
 \dot{\varphi}
\right)_0
+ 
\left(
\frac{\del \mathbf{L}^{ij}}{\del \dot{\bar{\varphi}}}
 \dot{\bar{\varphi}}
\right)_0
 -  \mathbf{L}^{ij} |_0.
\end{align}
For Lorentz invariant models, we can show
$\left(
\frac{\del \mathbf{L}^{0i}}{\del \dot{\varphi}}
 \dot{\varphi}
\right)_0
=
\left(
\frac{\del \mathbf{L}^{0i}}{\del \dot{\bar{\varphi}}}
 \dot{\bar{\varphi}}
\right)_0
= 0
$,
hence $\mathbf{M}^{0i}|_0 = 0$, 
in the case of temporal modulation but $\mathbf{M}^{00}|_0$ and
$\mathbf{M}^{ij}|_0$ are not equal to $\mathbf{L}^{00}|_0$,
$\mathbf{L}^{ij}|_0$, in general.
Since $\mathbf{L}^{00}|_0 = 0$, the kinetic term of the fluctuation
along the time-direction vanishes identically.
This distinguishes the temporal modulation from the spatial one.

\paragraph{Light-like modulation}
We define the light-like modulation as a vacuum characterized by the VEV:
\begin{align}
\langle 
\dot{\varphi}
\rangle \not= 0, \quad 
\langle \partial_1 \varphi \rangle \not= 0, \quad 
\langle \partial_2 \varphi \rangle = \langle \partial_3 \varphi \rangle
 = 0, \quad 
\del \varphi \cdot \del \varphi = \del \varphi \cdot \del \bar{\varphi}
 = 0.
\label{eq:ll}
\end{align}
Here $\del \varphi \cdot \del \varphi = \eta^{mn} \del_m \varphi \del_n
\varphi$ and so on. The metric is $\eta_{mn} = \mathrm{diag} (-1,1,1,1)$.
In order to look for the configuration \eqref{eq:ll}, we assume
\begin{align}
\dot{\varphi} = \pm \del_{1} \varphi.
\end{align}
If we demand the following conditions:
\begin{align}
& \mathbf{L}^{00}|_0 = \mathbf{L}^{0i}|_0 = \mathbf{L}^{ij}|_0 = 0,
\notag \\
& \left. \frac{\del \mathcal{L}}{\del (\del_i \varphi)} \right|_0 = 0,
\label{eq:conditions_temp_mod}
\end{align}
then, both the conditions \eqref{eq:eom}, \eqref{eq:energy_extremum}
for modulated vacua are satisfied.
However, these conditions are generally too strong and we will find
relaxed conditions in a concrete model of light-like modulation in
section \ref{sect:6th_model}.

\section{Generalized Nambu-Goldstone modes} \label{sect:NG_theorem}

In the previous section, we made a classification of modulated vacua
according to the spontaneous breaking of symmetries.
Along with the spontaneous symmetry breakings, we expect that NG
modes appear in the spectrum.
Particular emphasis is placed on the rotational symmetry.
Since the rotations in the $(x^1,x^2)$, $(x^1,x^3)$ planes are not
independent of the translation along the
$x^1$-direction \cite{Low:2001bw}, there 
are no independent NG modes associated with the spontaneous breaking of the
rotational symmetries. Therefore the only relevant part of our interest
is $U(1) \times P^{\hat{m}} \to [U(1) \times P^{\hat{m}}]_{\text{diag}}$ where $P^{\hat{m}}$ is space-like,
time-like or light-like. The NG modes associated with these symmetry
breakings are well-captured by the notion of the generalized NG modes in Ref.~\cite{Nitta:2017mgk}.
In this section, we discuss a generalization of the NG theorem
in our setup.
This provides a relation between zero modes of the generalized mass
matrix \eqref{eq:generalized_mass_matrix} and spontaneous symmetry breaking.
Most of the analysis in this section has been already discussed in
Ref.~\cite{Nitta:2017mgk}.
We therefore provide a brief sketch of the notion in the following.

We assume that the energy functional, $\mathcal{H}$, depends on the spacetime
derivative of scalar fields only : $\mathcal{H} (\varphi_m, \bar{\varphi}_m)$. 
Here we have defined $\varphi_m \equiv \del_m \varphi$,
$\bar{\varphi}_m \equiv \del_m \bar{\varphi}$.
Vacua are defined as extrema of $\mathcal{H}$ with respect to
$\varphi_m, \bar{\varphi}_m$:
\begin{align}
\frac{\partial \mathcal{H}}{\partial \varphi_m} = \frac{\partial
 \mathcal{H}}{\partial\bar{\varphi}_m} = 0.
\label{eq:extrema}
\end{align}
In these extrema, we assume $\varphi_m,\bar{\varphi}_m$ have the
following VEVs:
\begin{align}
\langle 0 | \varphi_m | 0 \rangle \equiv v_m = \delta^{\hat{m}} {}_m v, \qquad 
\langle 0 | \bar{\varphi}_m | 0 \rangle \equiv \bar{v}_m =
 \delta^{\hat{m}} {}_m \bar{v}.
\label{eq:VEVs}
\end{align}
Here $\hat{m}$ is a fixed spacetime direction.
The VEV $v_m$ can depend on the spacetime coordinate in general.
In that case, the vacuum spontaneously breaks spacetime symmetries.
Consider the fluctuations $\phi,\bar{\phi}$
around the VEV.
As we have discussed in the previous section, the expansion of the
Hamiltonian around the vacua results in ($\vec{\Phi}$ is defined
in eq.~\eqref{eq:generalized_vector})
\begin{align}
\mathcal{H} (v + \phi, \bar{v} + \bar{\phi})
 =& \ 
\mathcal{H} (v,\bar{v}) + 
 \frac{1}{2} \vec{\Phi}^{\dagger} \mathcal{M}|_0 \vec{\Phi} + \cdots.
\end{align}
In order that the vacua are local minima, we require that
$\mathcal{M}|_0$ is positive semi-definite.
The zero modes of $\mathcal{M}|_0$ correspond to flat directions.

Let us clarify the relation between the zero modes of $\mathcal{M}|_0$
and symmetries of the theory.
The fields $\varphi_m,\bar{\varphi}_m$ transform as 
\begin{align}
[i \mathcal{Q}^A, \varphi_m] = i (T^A \varphi)_m, \qquad 
[i \mathcal{Q}^A, \bar{\varphi}_m] = - i (T^A \bar{\varphi})_m.
\end{align}
Here $\mathcal{Q}^A$ are generators of the symmetry group and $T^A$ is a
Hermitian matrix representation of the generators.
As we mentioned before, a part of 
$\varphi_m$, $\bar{\varphi}_m$ develop VEVs \eqref{eq:VEVs}.
Then the spontaneous breaking of symmetries by modulations is defined 
if there exist operators $\varphi_{\hat{m}}$ with $v_{\hat{m}} \not= 0$ such that 
\begin{align}
\langle 0 | [i \mathcal{Q}^A, \varphi_{\hat{m}}] | 0 \rangle =  i (T^A v)_{\hat{m}}, \qquad 
\langle 0 | [i \mathcal{Q}^A, \bar{\varphi}_{\hat{m}}] | 0 \rangle = - i
 (T^A \bar{v})_{\hat{m}}.
\end{align}
If we define 
\begin{align}
(T^A \vec{v})_{\hat{m}} \equiv 
\left(
\begin{array}{c}
(T^A v)_{\hat{m}} \\
- (T^A \bar{v})_{\hat{m}}
\end{array}
\right),
\end{align}
then the symmetry corresponding to $T^{\hat{A}}$ with
$T^{\hat{A}} \vec{v} = 0$ is preserved in the vacuum while $T^{A'}$ such that 
$T^{A'} \vec{v} \not= 0$ is spontaneously broken.

The energy functional $\mathcal{H}$ is invariant under the following
transformation: 
\begin{align}
\varphi_m \to \varphi_m + i \varepsilon^A (T^A \varphi)_m, \qquad 
\bar{\varphi}_m \to \bar{\varphi}_m - i \varepsilon^A (T^A \bar{\varphi})_m,
\end{align}
where $\varepsilon^A$ is an infinitesimal parameter. 
Namely, we have 
\begin{align}
\delta^A \mathcal{H} = \frac{\partial \mathcal{H}}{\partial \varphi_n} (T^A \varphi)_n -
 \frac{\partial \mathcal{H}}{\partial\bar{\varphi}_n} (T^A\bar{\varphi})_n
 = 0.
\end{align}
By differentiating this relation with respect to
$\varphi_m,\bar{\varphi}_m$, we obtain
\begin{align}
& \frac{\partial^2 \mathcal{H}}{\partial \varphi_m \partial \varphi_n} (T^A \varphi)_n
+ \frac{\partial \mathcal{H}}{\partial \varphi_n} \frac{\partial}{\partial \varphi_m }
 (T^A \varphi)_n - \frac{\partial^2 \mathcal{H}}{\partial \varphi_m \partial \bar{\varphi}_n} (T^A \bar{\varphi})_n
 = 0, 
\notag \\
& \frac{\partial^2 \mathcal{H}}{\partial\bar{\varphi}_m \partial \varphi_n}
(T^A \varphi)_n - \frac{\partial^2 \mathcal{H}}{\partial\bar{\varphi}_m
 \partial\bar{\varphi}_n} (T^A \bar{\varphi})_n - \frac{\partial
 \mathcal{H}}{\partial\bar{\varphi}_n} \frac{\partial}{\partial
 \bar{\varphi}_m} (T^A \bar{\varphi})_n = 0.
\end{align}
If we consider these relations in the vacuum, the factor
$\partial \mathcal{H}/ \partial\varphi_m$ vanishes and we find
\begin{align}
\left(
\begin{array}{cc}
\left. \frac{\partial^2 \mathcal{H}}{\partial \bar{\varphi}_m \partial
 \varphi_n} \right|_0 
& 
\left. \frac{\partial^2 \mathcal{H}}{\partial \bar{\varphi}_m \partial
 \bar{\varphi}_n} \right|_0 
\\
\left. \frac{\partial^2 \mathcal{H}}{\partial \varphi_m \partial \varphi_n} \right|_0
&
\left. \frac{\partial^2 \mathcal{H}}{\partial \varphi_m \partial \bar{\varphi}_n} \right|_0
\end{array}
\right)
\left(
\begin{array}{c}
(T^A v)_n \\
- (T^A \bar{v})_n
\end{array}
\right) = 0.
\label{eq:vacuum_vector}
\end{align}
This results in the following relation:
\begin{align}
\mathbf{M}^{mn}|_0 (T^A \vec{v})_n = 0,
\label{eq:vacuum_vector2}
\end{align}
or equivalently
\begin{align}
\mathcal{M}|_0 (T^A \vec{V}) = 0.
\end{align}
Here we have defined $\langle 0| \vec{\Phi} |0 \rangle = \vec{V}$ where
$\vec{\Phi}$ is defined in eq.~\eqref{eq:generalized_vector}.
If $T^A \vec{V} \not= 0$, namely, $T^A = T^{A'}$ is a broken
generator, this is an eigenvector of the generalized mass 
matrix $\mathcal{M}|_0$ associated with zero eigenvalue\footnote{
Note that this relation never implies that all the zero modes of
$\mathcal{M}|_0$ are given by $T^A \vec{V}$ but the inverse is true.
Namely if $T^A \vec{V} \not= 0$ exists, it should be a zero mode of $\mathcal{M}|_0$.
}.
We call these zero modes $T^A \vec{V}$ the generalized NG modes.
Note that since only parts of the component in $\vec{V}$ are nonzero,
{\it i.e.}, 
\begin{align}
\vec{V} = 
\left(
\begin{array}{c}
0 \\
\vdots \\
(T^A \vec{v})_{\hat{m}} \\
\vdots
\\
0
\end{array}
\right),
\end{align}
only the subsector that corresponds to the nonzero VEV directions
$\hat{m}$ are relevant to specify the zero modes.

\section{Higher derivative scalar model}
\label{sect:6th_model}

In this section, we demonstrate the existence of modulated vacua by
focusing on a concrete model with Lorentz invariance, and study
fluctuations in each case according to the classification of spatial,
temporal and light-like modulation.

\subsection{Model}

The model that we will consider here, is based on the global stability
considerations presented in Ref.~\cite{Nitta:2017mgk}, which we will
briefly review.
We consider a Lagrangian density which only depends on $\del_m\varphi$
and its complex conjugate,
$\mathcal{L}(\del_m\varphi,\del_m\bar{\varphi})$ and it contains a
finite power of the derivative term,
$|\del\varphi|^2=\del_m\varphi\del^m\bar{\varphi}$,
hence the highest power of the derivative takes the form:
\begin{equation}
\mathcal{L} \supset \mp|\del\varphi|^{2N}
= \mp\left(-|\dot{\varphi}|^2 + |\nabla\varphi|^2\right)^N,
\end{equation}
with $N\in\mathbb{Z}_{>0}$ being a positive integer.
The spacetime index $m$ has been contracted by the inverse
metric $\eta^{mn}={\rm diag}(-1,1,1,1)$ yielding the relative minus
sign in the last expression.
$\dot{\varphi}$ is the time derivative of $\varphi$ and
$\nabla\varphi$ is the spatial gradient vector of $\varphi$. 
Let us now construct the Hamiltonian by performing a Legendre
transformation
\begin{equation}
\pi = \frac{\del\mathcal{L}}{\del\dot{\varphi}}
\supset \mp(-1)^NN|\del\varphi|^{2N-2}\dot{\bar\varphi},
\end{equation}
yielding
\begin{equation}
\mathcal{H} = \pi\dot{\varphi} + \bar{\pi}\dot{\bar\varphi}
- \mathcal{L}
\supset \mp(-1)^N(2N-1)|\dot\varphi|^{2N} \pm |\nabla\varphi|^{2N}.
\end{equation}
The global stability of the vacuum in this model is given provided
that there is no runaway direction.
Considering first the limit of $|\nabla\varphi|^2\to\infty$, we can
conclude that we must choose the upper sign.
Similarly we consider the limit of $|\dot\varphi|^2\to\infty$. Since
we have chosen the upper sign, we see from the above Hamiltonian that
the largest power of the derivative terms, $N$, must be odd. 
$N=1$ is just the canonical kinetic term.
Hence, the simplest nontrivial example with global stability is
$N=3$. 

We will here consider a higher-derivative scalar field model of the
6th order, corresponding to $N=3$. 
The most general Lagrangian density of the
Lorentz-invariant scalar field theory with $N=3$ that
is compatible with the symmetries \eqref{eq:symmetries} and the
absence of the Ostrogradsky instability is given by 
\begin{align}
\mathcal{L} 
=& \ 
- k \del \varphi \cdot \del \bar{\varphi} 
+ \lambda (\del \varphi \cdot \del \varphi) (\del \bar{\varphi} \cdot
 \del \bar{\varphi}) + \mu (\del \varphi \cdot \del \bar{\varphi})^2
\notag \\
& \ + \alpha (\del \varphi \cdot \del \bar{\varphi}) (\del \varphi \cdot
 \del \varphi) (\del \bar{\varphi} \cdot \del \bar{\varphi})
+ \beta (\del \varphi \cdot \del \bar{\varphi})^3.
\label{eq:6th_bosonic}
\end{align}
The first term corresponds to the canonical kinetic term and the
other terms are higher-derivative corrections.
The parameters $k, \lambda, \mu, \alpha, \beta$ are constants.
When $\beta = \mu = 0$, the model reduces to the one studied in
Ref.~\cite{Nitta:2017mgk}\footnote{This limit is also the bosonic part
of the supersymmetric model studied in Ref.~\cite{Nitta:2017yuf}}.
The energy density is 
\begin{align}
\mathcal{H} =& \ 
2 k |\dot{\varphi}|^2 - 4 \mu |\dot{\varphi}|^2 (\del \varphi \cdot \del
 \bar{\varphi})
- 2 \alpha |\dot{\varphi}|^2 (\del \varphi \cdot \del \varphi) (\del
 \bar{\varphi} \cdot \del \bar{\varphi}) 
\notag \\
& 
- 6 \beta |\dot{\varphi}|^2
 (\del \varphi \cdot \del \bar{\varphi})^2 - 2 \lambda \dot{\varphi}^2
 (\del \bar{\varphi} \cdot \del \bar{\varphi}) - 2 \lambda
 \dot{\bar{\varphi}} (\del \varphi \cdot \del \varphi)
\notag \\
&
- 2 \alpha \dot{\varphi}^2 (\del \varphi \cdot \del \bar{\varphi}) (\del
 \bar{\varphi} \cdot \del \bar{\varphi}) - 2 \alpha \dot{\bar{\varphi}}
 (\del \varphi \cdot \del \bar{\varphi}) (\del \varphi \cdot \del
 \varphi)
\notag \\
&
+ k (\del \varphi \cdot \del \bar{\varphi})
- \lambda (\del \varphi \cdot \del \varphi) (\del \bar{\varphi} \cdot
 \del \bar{\varphi}) - \mu (\del \varphi \cdot \del \bar{\varphi})^2 
\notag \\
& 
- \alpha (\del \varphi \cdot \del \bar{\varphi}) (\del \varphi \cdot
 \del \varphi) (\del \bar{\varphi} \cdot \del \bar{\varphi})
- \beta (\del \varphi \cdot \del \bar{\varphi})^3.
\end{align}

In order to examine the conditions \eqref{eq:eom},
\eqref{eq:energy_extremum} for modulated vacua, we evaluate 
the matrices $\mathbf{L}^{mn}$.
For the model \eqref{eq:6th_bosonic}, we have 
\begin{align}
\mathbf{L}^{mn} &=
\begin{pmatrix}
\mathbf{L}^{mn}_{\bar{\varphi}\varphi} & \mathbf{L}^{mn}_{\bar{\varphi}\bar{\varphi}}\\
\mathbf{L}^{mn}_{\varphi\varphi} & \mathbf{L}^{mn}_{\varphi\bar{\varphi}}
\end{pmatrix},
\end{align}
where each component is calculated as 
\begin{align}
\mathbf{L}^{mn}_{\bar{\varphi}\varphi} &\equiv
- k \eta^{mn} + \alpha \eta^{mn} (\del \varphi \cdot \del \varphi)
 (\del \bar{\varphi} \cdot \del \bar{\varphi}) 
+ 2 \alpha \del^m \bar{\varphi} \del^n \bar{\varphi} (\del \varphi \cdot \del \varphi) 
+ 2 \alpha \del^m \varphi \del^n \varphi
 (\del \bar{\varphi} \cdot \del \bar{\varphi}) \non
&\phantom{=\ }
+ 4 (\lambda + \alpha \del \varphi \cdot \del \bar{\varphi}) \del^m
\bar{\varphi} \del^n \varphi
+ 2 \mu 
\left\{
\del^m \varphi \del^n \bar{\varphi} + \eta^{mn} (\del \varphi \cdot \del
\bar{\varphi})
\right\} \non
&\phantom{=\ }
+ 3 \beta (\del \varphi \cdot \del \bar{\varphi}) 
\left\{
2 \del^m \varphi \del^n \bar{\varphi} + \eta^{mn} (\del \varphi \cdot
\del \bar{\varphi})
\right\},\\
\mathbf{L}^{mn}_{\bar{\varphi}\bar{\varphi}} &\equiv
2 \alpha (\del \varphi \cdot \del \varphi)
(\del^m \bar{\varphi} \del^n \varphi + \del^m \varphi \del^n
\bar{\varphi})
+ 2 \eta^{mn} (\lambda + \alpha \del \varphi \cdot \del \bar{\varphi})
(\del \varphi \cdot \del \varphi)
+ 2 \mu \del^m \varphi \del^n \varphi \non
&\phantom{=\ }
+ 6 \beta (\del \varphi \cdot \del \bar{\varphi}) \del^m \varphi \del^n \varphi,\\
\mathbf{L}^{mn}_{\varphi\varphi} &\equiv
2 \alpha (\del \bar{\varphi} \cdot \del \bar{\varphi})
(\del^m \varphi \del^n \bar{\varphi} + \del^m \bar{\varphi} \del^n
\varphi)
+ 2 \eta^{mn} (\lambda + \alpha \del \varphi \cdot \del \bar{\varphi})
(\del \bar{\varphi} \cdot \del \bar{\varphi})
+ 2 \mu \del^m \bar{\varphi} \del^n \bar{\varphi} \non
&\phantom{=\ }
+ 6 \beta (\del \varphi \cdot \del \bar{\varphi}) \del^m \bar{\varphi}
\del^n \bar{\varphi}, \\
\mathbf{L}^{mn}_{\varphi\bar{\varphi}} &\equiv 
- k \eta^{mn} + \alpha \eta^{mn} (\del \varphi \cdot \del \varphi) (\del
 \bar{\varphi} \cdot \del \bar{\varphi})
+ 2 \alpha \del^m \varphi \del^n \varphi (\del \bar{\varphi} \cdot
 \del \bar{\varphi}) + 2 \alpha \del^m \bar{\varphi} \del^n
 \bar{\varphi} (\del \varphi \cdot \del \varphi) \non
&\phantom{=\ }
+ 4 (\lambda + \alpha \del \varphi \cdot \del \bar{\varphi}) \del^m
 \varphi \del^n \bar{\varphi}
+ 2 \mu 
\left\{
\del^m \bar{\varphi} \del^n \varphi + \eta^{mn} (\del \varphi \cdot \del \bar{\varphi})
\right\} \non
&\phantom{=\ }
+ 3 \beta (\del \varphi \cdot \del \bar{\varphi}) 
\left\{
2 \del^m \bar{\varphi} \del^n \varphi + \eta^{mn} (\del \varphi \cdot
\del \bar{\varphi})
\right\}.
\end{align}
Then, it is straightforward to calculate the other relevant
quantities, such as $\frac{\del\mathbf{L}^{mn}}{\del\dot{\varphi}}$ and its
complex conjugate:
\begin{align}
\frac{\del\mathbf{L}^{mn}_{\bar{\varphi}\varphi}}{\del\dot{\varphi}}
&= -2 \alpha \eta^{mn} \dot{\varphi} (\del
 \bar{\varphi} \cdot \del \bar{\varphi}) 
-4 \alpha \dot{\varphi} \del^m \bar{\varphi} \del^n \bar{\varphi}
+ 2 \alpha (\del \bar{\varphi} \cdot \del \bar{\varphi})
(\eta^{m0} \del^n \varphi + \del^m \varphi \eta^{n0}) \non
&\phantom{=\ }
- 4 \alpha \dot{\bar{\varphi}} \del^m
\bar{\varphi} \del^n \varphi
+ 4 (\lambda + \alpha \del \varphi \cdot \del \bar{\varphi}) 
\del^m \bar{\varphi} \eta^{n0}
+ 2 \mu 
\left(
\eta^{m0} \del^n \bar{\varphi} - \eta^{mn} \dot{\bar{\varphi}}
\right) \non
&\phantom{=\ }
- 3 \beta \dot{\bar{\varphi}} 
\left\{
2 \del^m \varphi \del^n \bar{\varphi} + \eta^{mn} (\del \varphi \cdot
\del \bar{\varphi})
\right\}
+ 3 \beta (\del \varphi \cdot \del \bar{\varphi}) 
\left\{
2 \eta^{m0} \del^n \bar{\varphi} - \eta^{mn} \dot{\bar{\varphi}}
\right\},\\
\frac{\del\mathbf{L}^{mn}_{\bar{\varphi}\bar{\varphi}}}{\del\dot{\varphi}}
&= -4 \alpha \dot{\varphi}
(\del^m \bar{\varphi} \del^n \varphi + \del^m \varphi \del^n
\bar{\varphi})
+ 2 \alpha (\del \varphi \cdot \del \bar{\varphi})
(\del^m \bar{\varphi} \eta^{n0} + \eta^{m0} \del^n \bar{\varphi})
- 2 \alpha \eta^{mn} \dot{\bar{\varphi}} \del \varphi \cdot \del
 \bar{\varphi} \non
&\phantom{=\ }
- 4 \eta^{mn} \dot{\varphi} (\lambda + \alpha \del \varphi \cdot \del
\bar{\varphi})
+ 2 \mu (\eta^{m0} \del^n \varphi + \eta^{n0} \del^m \varphi)
- 6 \beta \dot{\bar{\varphi}} \del^m \varphi \del^n \varphi \non
&\phantom{=\ }
+ 6 \beta (\del \varphi \cdot \del \bar{\varphi})
(\eta^{m0} \del^n \varphi + \eta^{n0} \del^m \varphi),\\
\frac{\del\mathbf{L}^{mn}_{\varphi\varphi}}{\del\dot{\varphi}}
&= 2 \alpha (\del \bar{\varphi} \cdot \del \bar{\varphi})
(\eta^{m0} \del^n \bar{\varphi} + \del^m \bar{\varphi} \eta^{n0})
-2 \alpha \eta^{mn} \dot{\bar{\varphi}} (\del \bar{\varphi} \cdot \del
 \bar{\varphi})
 - 6 \beta \dot{\bar{\varphi}} \del^m \bar{\varphi} \del^n
 \bar{\varphi},
\\
\frac{\del\mathbf{L}^{mn}_{\varphi\bar{\varphi}}}{\del\dot{\varphi}}
&= -2 \alpha \eta^{mn} \dot{\varphi} (\del \bar{\varphi} \cdot \del \bar{\varphi})
-4 \alpha \dot{\varphi} \del^m \bar{\varphi} \del^n \bar{\varphi}
+ 2 \alpha (\del \bar{\varphi} \cdot \del \bar{\varphi}) (\eta^{m0}
\del^n \varphi + \del^m \varphi \eta^{n0}) \non
&\phantom{=\ }
- 4 \alpha \dot{\bar{\varphi}} \del^m \varphi \del^n \bar{\varphi} + 4
 (\lambda + \alpha \del \varphi \cdot \del \bar{\varphi}) \eta^{m0}
 \del^n \bar{\varphi}
+ 2 \mu 
\left(
\eta^{n0} \del^m \bar{\varphi} - \eta^{mn} \dot{\bar{\varphi}}
\right) \non
&\phantom{=\ }
- 3 \beta \dot{\bar{\varphi}} 
\left\{
2 \del^m \bar{\varphi} \del^n \varphi + \eta^{mn} (\del \varphi \cdot
\del \bar{\varphi})
\right\}
+ 6 \beta (\del \varphi \cdot \del \bar{\varphi}) (2 \eta^{n0} \del^m
 \bar{\varphi} - \eta^{mn} \dot{\bar{\varphi}}), 
\end{align}
\begin{align}
\frac{\del\mathbf{L}^{mn}_{\bar{\varphi}\varphi}}{\del\dot{\bar\varphi}}
&= -2 \alpha \eta^{mn} \dot{\bar{\varphi}} (\del \varphi \cdot \del
 \varphi)
+ 2 \alpha (\del \varphi \cdot \del \varphi) (\eta^{m0} \del^n
 \bar{\varphi} + \del^m \bar{\varphi} \eta^{n0})
- 4 \alpha \dot{\bar{\varphi}} \del^m \varphi \del^n \varphi \non
&\phantom{=\ }
- 4 \alpha \dot{\varphi} \del^m \bar{\varphi} \del^n \varphi 
+ 4 (\lambda + \alpha \del \varphi \cdot \del \bar{\varphi}) \del^n
\varphi \eta^{m0}
+ 2 \mu 
\left(
\eta^{n0} \del^m \varphi - \eta^{mn} \dot{\varphi}
\right) \non
&\phantom{=\ }
- 3 \beta \dot{\varphi} 
\left\{
2 \del^m \varphi \del^n \bar{\varphi} + \eta^{mn} (\del \varphi \cdot
\del \bar{\varphi})
\right\}
+ 3 \beta (\del \varphi \cdot \del \bar{\varphi}) 
\left(
2 \eta^{n0} \del^m \varphi - \eta^{mn} \dot{\varphi}
\right),
\\
%
\frac{\del\mathbf{L}^{mn}_{\bar{\varphi}\bar{\varphi}}}{\del\dot{\bar\varphi}}
&= 2 \alpha (\del \varphi \cdot \del \varphi) (\eta^{m0} \del^n \varphi +
 \del^m \varphi \eta^{n0})
- 2 \alpha \eta^{mn} \dot{\varphi} (\del \varphi \cdot \del \varphi)
- 6 \beta \dot{\varphi} \del^m \varphi \del^n \varphi,\\
\frac{\del\mathbf{L}^{mn}_{\varphi\varphi}}{\del\dot{\bar\varphi}}
&= - 4 \alpha \dot{\bar{\varphi}} (\del^m \varphi \del^n \bar{\varphi} +
 \del^m \bar{\varphi} \del^n \varphi)
+ 2 \alpha (\del \bar{\varphi} \cdot \del \bar{\varphi}) (\del^m \varphi
 \eta^{n0} + \eta^{m0} \del^n \varphi) \non
&\phantom{=\ }
-2 \alpha \eta^{mn} \dot{\varphi} (\del \bar{\varphi} \cdot \del \bar{\varphi})
- 4 \eta^{mn} \dot{\bar{\varphi}} (\lambda + \alpha \del \varphi \cdot
 \del \bar{\varphi})
+ 2 \mu 
\left(
\eta^{m0} \del^n \bar{\varphi} + \eta^{n0} \del^m \bar{\varphi}
\right) \non
&\phantom{=\ }
- 6 \beta \dot{\varphi} \del^m \bar{\varphi} \del^n \bar{\varphi}
+ 6 \beta (\del \varphi \cdot \del \bar{\varphi}) 
\left(
\eta^{m0} \del^n \bar{\varphi} + \eta^{n0} \del^m \bar{\varphi}
\right),\\
\frac{\del\mathbf{L}^{mn}_{\varphi\bar{\varphi}}}{\del\dot{\bar\varphi}}
&= - 2 \alpha \eta^{mn} \dot{\bar{\varphi}} \del \varphi \cdot \del \varphi 
+ 2 \alpha (\del \varphi \cdot \del \varphi) (\eta^{m0} \del^n
 \bar{\varphi} + \del^m \bar{\varphi} \eta^{n0})
- 4 \alpha \dot{\bar{\varphi}} \del^m \varphi \del^n \varphi \non
&\phantom{=\ }
- 4 \alpha \dot{\varphi} \del^m \varphi \del^n \bar{\varphi}
+ 4 (\lambda + \alpha \del \varphi \cdot \del \bar{\varphi}) \del^m
\varphi \eta^{n0}
+ 2 \mu 
\left(
\eta^{m0} \del^n \varphi - \eta^{mn} \dot{\varphi}
\right) \non
&\phantom{=\ }
- 3 \beta \dot{\varphi} 
\left\{
2 \del^m \bar{\varphi} \del^n \varphi + \eta^{mn} (\del \varphi \cdot
\del \bar{\varphi})
\right\}
+ 3 \beta (\del \varphi \cdot \del \bar{\varphi}) 
\left(
2 \eta^{m0} \del^n \varphi - \eta^{mn} \dot{\varphi}
\right).
\end{align}
We also have
\begin{align}
\frac{\del \mathcal{L}}{\del (\del_i \varphi)} =& \ 
\left\{
\frac{}{}
-k + \alpha (\del \varphi \cdot \del \varphi) (\del \bar{\varphi} \cdot
 \del \bar{\varphi}) 
+ 2 \mu (\del \varphi \cdot \del \bar{\varphi})
+ 3 \beta (\del \varphi \cdot \del \bar{\varphi})^2
\right\} \del^i \bar{\varphi} 
\notag \\
& \ 
+ 2 (\lambda + \alpha \del \varphi \cdot \del \bar{\varphi}) (\del
 \bar{\varphi} \cdot \del \bar{\varphi}) \del^i \varphi,
\notag \\
\frac{\del \mathcal{L}}{\del (\del_i \bar{\varphi})} =& \  
\left\{
\frac{}{}
-k + \alpha (\del \varphi \cdot \del \varphi) (\del \bar{\varphi} \cdot
 \del \bar{\varphi}) 
+ 2 \mu (\del \varphi \cdot \del \bar{\varphi})
+ 3 \beta (\del \varphi \cdot \del \bar{\varphi})^2
\right\} \del^i \varphi
\notag \\
& \ 
+ 2 (\lambda + \alpha \del \varphi \cdot \del \bar{\varphi}) (\del
 \varphi \cdot \del \varphi) \del^i \bar{\varphi}.
\end{align}
Now that we have all the ingredients to examine the conditions
\eqref{eq:eom}, \eqref{eq:energy_extremum}, 
we will study the spatial, temporal and light-like modulations in
turn. 

\subsection{Spatial modulation}
We first consider the spatially modulated vacuum in the model.
As we will show in the following, 
the results in this subsection are essentially the same with the ones in
Ref.~\cite{Nitta:2017mgk}, 
but Lagrangian here is more general.
In order to make the paper be self-contained, we rederive the
results in the more general setup \eqref{eq:6th_bosonic}.

For the spatial modulation, we employ the Ansatz
\begin{align}
\langle \varphi \rangle = \varphi_0 e^{i c x^1},
\label{eq:ansatz_sp}
\end{align}
where $\varphi_0,c$ are real constants and 
the VEV is of the FF-type in the $x^1$-direction.
Following the general procedure discussed in section
\ref{subsect:classification_mod_vac}, we examine the spatially
modulated vacuum. 
The Ansatz implies
\begin{align}
\langle \dot{\varphi} \rangle = \langle \del_2 \varphi \rangle = \langle
 \del_3 \varphi \rangle = 0, \qquad \langle \partial_1 \varphi  \rangle
 = i c \varphi_0 e^{i c x^1} \not= 0.
\end{align}
Then, we have
\begin{align}
\frac{\del \mathcal{L}}{\del (\del_i \varphi)} =& \
\left[
-k + \alpha |\del_1 \varphi|^4 + 2 \mu |\del_1 \varphi|^2 + 3 \beta
 |\del_1 \varphi|^4
\right] \delta^i_1 \del_1 \bar{\varphi}
+ 2 (\lambda + \alpha |\del_1 \varphi|^2) (\del_1 \bar{\varphi})^2
 \delta^i_1 \del_1 \varphi .
\end{align}
We first study the condition \eqref{eq:energy_extremum}.
As discussed in section \ref{subsect:classification_mod_vac},
the nontrivial conditions boil down to
$\frac{\del \mathcal{L}}{\del (\del_i \varphi)} = 0$.
The conditions for $i=2,3$ in eq.~\eqref{eq:sp1} are automatically satisfied.
On the other hand, for $i = 1$, we have
\begin{align}
0 = 
\frac{\del \mathcal{L}}{\del (\del_1 \varphi)} =& \ 
\left\{
-k + \alpha |\del \varphi|^4 + 2 \mu |\del_1 \varphi|^2 + 3 \beta |\del
 \varphi|^4 
\right\} \del_1 \bar{\varphi}
+ 2 (\lambda + \alpha |\del \varphi|^2) |\del_1 \varphi|^2 \del_1
 \bar{\varphi}
\notag \\
= & \ 
\del_1 \bar{\varphi}
\left[
\frac{}{}
- k + 2 (\mu + \lambda) |\del_1 \varphi|^2 + 3 (\alpha +  \beta )
 |\del_1 \varphi|^4
\right].
\label{eq:sp_amp}
\end{align}
This determines the amplitude of the VEV:
\begin{align}
|\del_1 \varphi|^2 = |c \varphi_0|^2 = 
\frac{- (\lambda + \mu) \pm \sqrt{(\lambda + \mu)^2 + 3 k (\alpha +
 \beta)}}{3 (\alpha + \beta)}.
\label{eq:sp_mod_sol}
\end{align}
Next, we analyze the other condition, {\it i.e.}~eq.~\eqref{eq:eom}.
The matrices $\mathbf{L}^{mn}$ for the spatial modulation are calculated as
\begin{align}
\mathbf{L}^{00} =& \ 
\left(
\begin{array}{c|c}
k - 2 \beta |\del_1 \varphi|^2 - (\alpha + 3 \beta) |\del_1 \varphi|^4
&
-2 (\del_1 \varphi)^2 (\lambda + \alpha |\del_1 \varphi|^2)
 \\
\hline
- 2 (\del_1 \bar{\varphi})^2 (\lambda + \alpha |\del_1 \varphi|^2)
&
k - 2 \beta |\del_1 \varphi|^2 - (\alpha + 3 \beta) |\del_1 \varphi|^4
\end{array}
\right),\\
\mathbf{L}^{11} =& \ 
\left(
\begin{array}{c|c}
- k + 4 (\lambda + \mu) |\del_1 \varphi|^2 + 9 (\alpha + \beta) |\del_1 \varphi|^4
&
2 (\del_1 \varphi)^2 
\left\{
(\lambda + \mu) + 3 (\alpha + \beta) |\del_1 \varphi|^2
\right\}
\\
\hline
2 (\del_1 \bar{\varphi})^2 
\left\{
(\lambda + \mu) + 3 (\alpha + \beta) |\del_1 \varphi|^2
\right\}
&
- k + 4 (\lambda + \mu) |\del_1 \varphi|^2 + 9 (\alpha + \beta) |\del_1 \varphi|^4
\end{array}
\right), 
\notag \\
\mathbf{L}^{22} =& \ \mathbf{L}^{33} = 
\left(
\begin{array}{c|c}
-k + 2 \mu |\del_1 \varphi|^2 + (\alpha + 3 \beta) |\del_1 \varphi|^4
&
2 (\del_1 \varphi)^2 (\lambda + \alpha |\del_1 \varphi|^2)
 \\
\hline 
2 (\del_1 \bar{\varphi})^2 (\lambda + \alpha |\del_1 \varphi|^2)
&
-k + 2 \mu |\del_1 \varphi|^2 + (\alpha + 3 \beta) |\del_1 \varphi|^4
\end{array}
\right), \notag
\end{align}
and the others vanish.
One immediately finds that the only nontrivial condition comes from the
$(i,j) = (1,1)$ component of the third term in eq.~\eqref{eq:eom}. 
However, the term should vanish from the general discussion. 
Let us confirm this fact. Although $\mathbf{L}^{11} \not=0$, we have
\begin{align}
\mathbf{L}^{11} 
\left(
\begin{array}{c}
\del_1^2 \varphi
 \\
\del_1^2 \bar{\varphi}
\end{array}
\right) =& \ 
ic 
\left(
\begin{array}{c}
\del_1 \varphi
\left[
-k + 2 (\lambda + \mu) |\del_1 \varphi|^2 + 3 (\alpha + \beta) |\del_1 \varphi|^4
\right]
 \\
- \del_1 \bar{\varphi} 
\left[
-k + 2 (\lambda + \mu) |\del_1 \varphi|^2 + 3 (\alpha + \beta) |\del_1 \varphi|^4
\right]
\end{array}
\right).
\end{align} 
This vanishes due to the condition 
\eqref{eq:sp_amp} as expected.

\paragraph{Global aspects of the potential and stability of the vacuum}
Although we have confirmed that the spatially modulated configuration
\eqref{eq:ansatz_sp} is a solution to the equation of motion and the
energy extremum condition, it is still unclear whether the configuration
is a (global or local) minimum of the energy functional.
As we will show below, this holds true in a specific region of the
parameters $k,\alpha,\beta,\lambda,\mu$.

\begin{figure}[t]
\begin{center}
\includegraphics[scale=.38]{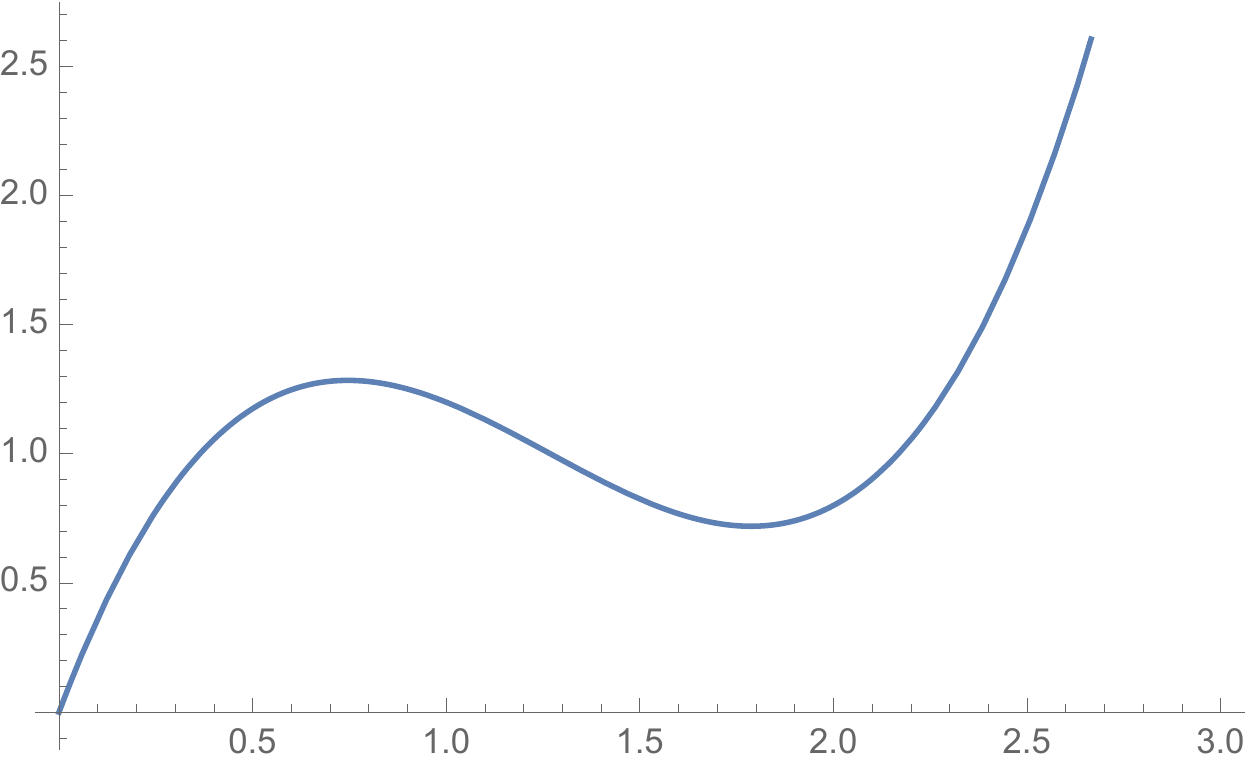}
\includegraphics[scale=.38]{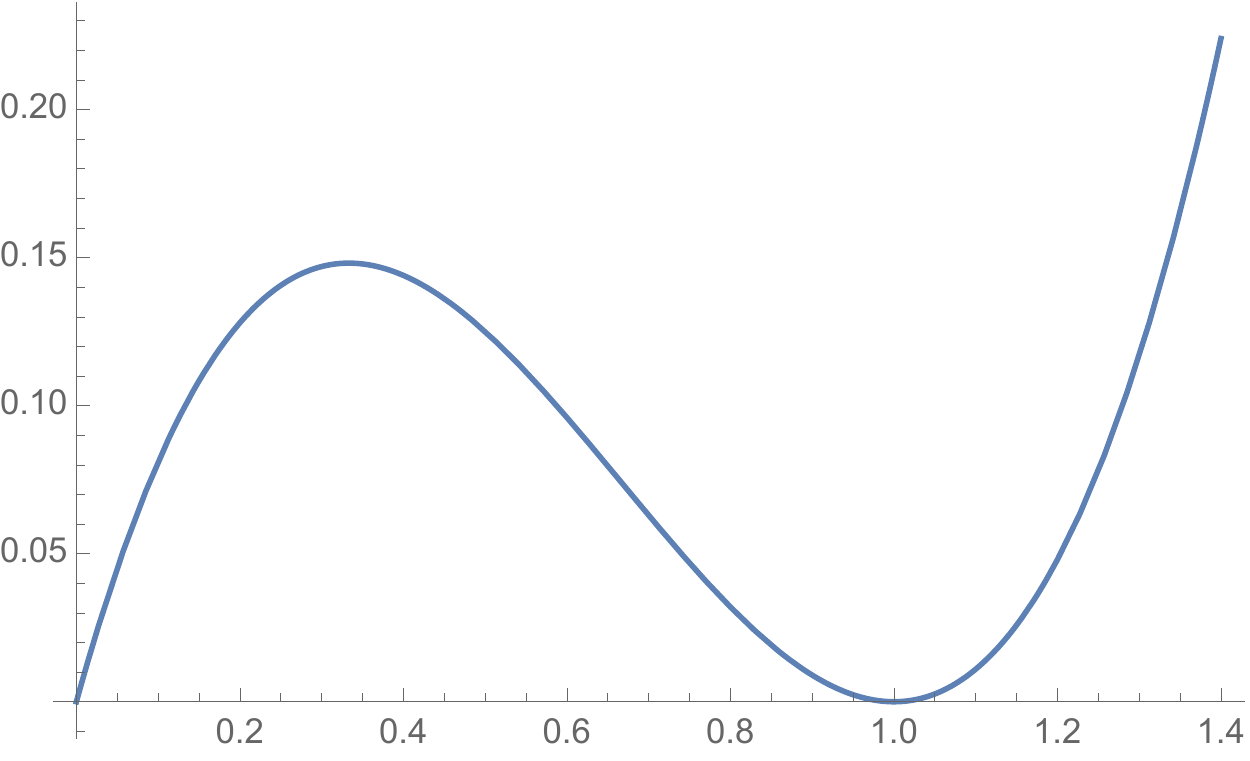}
\includegraphics[scale=.38]{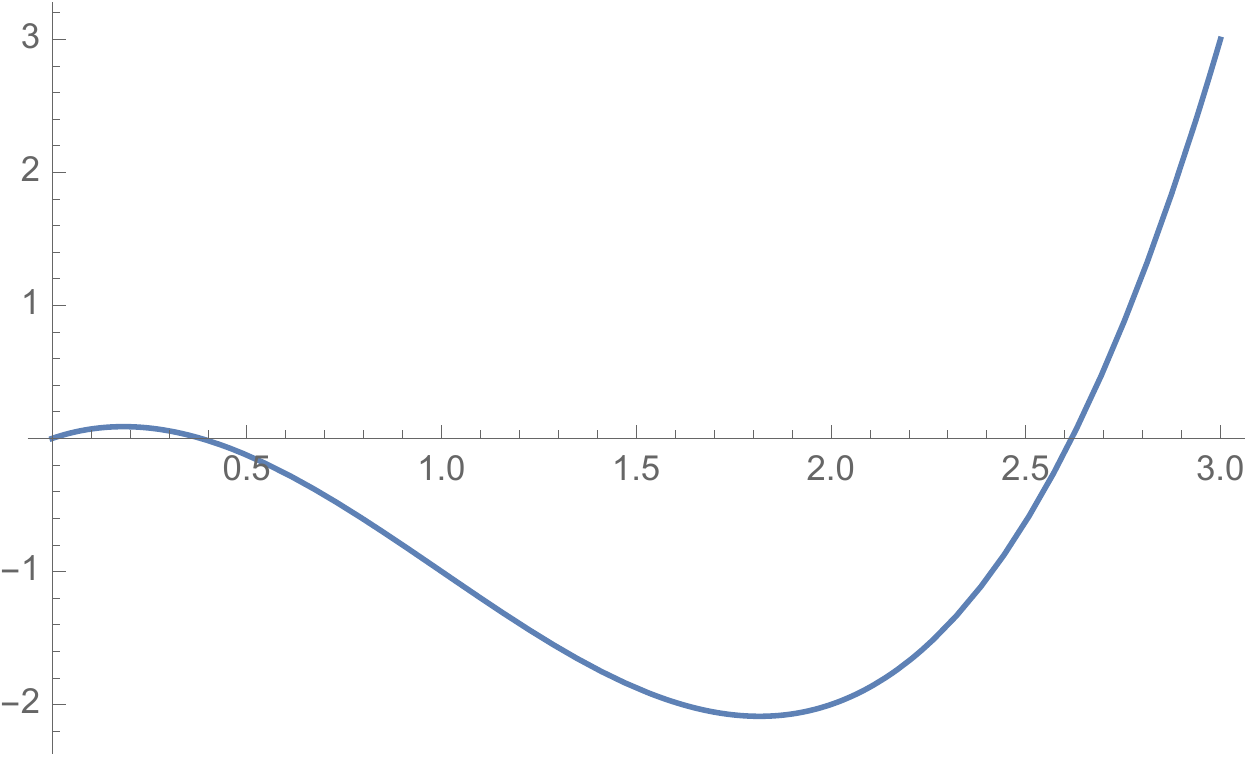}
\end{center}
\caption{The energy density as a function of $X=|\del_1 \varphi|^2$.
The vertical axis represents the energy density while the horizontal
 axis represents $X$.
The parameters are, left (meta-stable): $k=4$, $\lambda + \mu = 3.8$,
 $\alpha + \beta = -1$, middle (degenerate): $k=1$, $\lambda + \mu = 2$, $\alpha + \beta = -1$,
right (global): $k=1$, $\lambda + \mu = 3$, $\alpha + \beta = -1$.}
\label{fig:sp_pot}
\end{figure}

We first study the global stability in the $\del_1 \varphi$-direction.
The energy density for the Ansatz \eqref{eq:ansatz_sp} is
\begin{align}
\mathcal{E}_{\text{sp}} = 
k |\del_1 \varphi|^2 - (\lambda + \mu)
 |\del_1 \varphi|^4 - (\alpha + \beta) |\del_1 \varphi|^6.
\end{align}
This is a function of $X \equiv |\del_1 \varphi|^2 \ge 0$: $y = k X - (\lambda + \mu) X^2 -
(\alpha + \beta) X^3$.
From this expression the condition for the existence of a local minimum
is determined by the discriminant condition of $y' = k - 2 (\lambda + \mu) X - 3 (\alpha +
\beta) X^2 = 0$:
\begin{align}
\mathbf{D} = 4 \left\{
(\lambda + \mu)^2 + 3 k (\alpha + \beta)
\right\} > 0. 
\label{eq:sp_discriminant}
\end{align}
This is the condition that the energy has a local minimum
at $X \not= 0$.
When these conditions are satisfied, one finds that the solution \eqref{eq:sp_mod_sol}
is a local minimum of the energy.
Since the energy is given in the form
\begin{align}
\mathcal{E}_{\text{sp}} = X 
\left[
\frac{}{}
-( \alpha + \beta) X^2 - (\lambda + \mu) X + k 
\right], \qquad X \equiv |\del_1 \varphi|^2,
\end{align}
one finds that if $f (X) \equiv -( \alpha + \beta) X^2 - (\lambda + \mu) X + k = 0$
has no solution, then, the local minimum \eqref{eq:sp_mod_sol} is
meta-stable. If $f (X) = 0$ has one solution, then the vacuum
\eqref{eq:sp_mod_sol} is degenerate with the trivial vacuum $X=0$.
If $f (X) = 0$ has two solutions, then \eqref{eq:sp_mod_sol} becomes the
global minimum. In summary, we have the following conditions for the
global structures of the ``energy potential'' $\mathcal{E}_{\text{sp}}$ (see Fig.~\ref{fig:sp_pot}):
\begin{align}
 \text{meta-stable} \ :& \ 
(\lambda + \mu)^2 + 4 k (\alpha + \beta) < 0,
\notag \\
 \text{degenerate} \ :& \
(\lambda + \mu)^2 + 4 k (\alpha + \beta) = 0,
\notag \\
 \text{global} \ :& \ 
(\lambda + \mu)^2 + 4 k (\alpha + \beta) > 0.
\end{align}

We now discuss the stability of the vacuum including the
$\dot{\varphi}, \del_2 \varphi, \del_3 \varphi$-directions.
The local stability is guaranteed by the condition $\mathcal{M}|_0 \ge 0$.
As discussed in section \ref{subsect:classification_mod_vac}, 
since the ``off-diagonal parts'' of $\mathbf{L}^{mn}$ vanish and
$\dot{\varphi} = \dot{\bar{\varphi}} = 0$, 
the matrices $\mathbf{M}^{mn}$ are simply given by 
eq.~\eqref{eq:sp_Lagrangian_matrices} 
and the generalized mass matrix is
\begin{align}
\mathcal{M}|_0 = 
\left(
\begin{array}{cccc}
\mathbf{M}^{00}|_0 & & &  \\
& \mathbf{M}^{11}|_0 & & \\
& & \mathbf{M}^{22}|_0 & \\
& & & \mathbf{M}^{33}|_0
\end{array}
\right),
\end{align}
which is totally block diagonal. 
One can easily find eigenvalues of each sector.
The eigenvalues of $\mathbf{M}^{00}|_0$ are 
\begin{align}
A_1 =& \ 
\frac{2 (\beta -\mu ) \Upsilon}{3 (\alpha +\beta )},
\qquad
A_2 = \ 
\frac{12 \alpha  k (\alpha +\beta )
-2 \big(3 \beta  (-\beta +2 \lambda +\mu)-\alpha  (3 \beta -2 \lambda +\mu )\big) \Upsilon}{9 (\alpha +\beta )^2},
\end{align}
where we have defined
\begin{equation}
\Upsilon\equiv \sqrt{3 k (\alpha +\beta )+(\lambda +\mu
 )^2}+\lambda +\mu,
\end{equation}
and the eigenvalues of $\mathbf{M}^{11}|_0$ are 
\begin{align}
B_1 =& \ 0,
\qquad
B_2 = \ 
-\frac{4 (\lambda +\mu ) \Upsilon}{3 (\alpha +\beta )}-4 k.
\end{align}
The eigenvalues of $\mathbf{M}^{22}|_0$ and  $\mathbf{M}^{33}|_0$ are 
\begin{align}
C_1 =& \ 0,
\qquad
C_2 = \ 
\frac{12 \alpha  k (\alpha +\beta )-4 \big(\alpha  (\lambda -2 \mu )+3 \beta
 \lambda \big) \Upsilon}{9 (\alpha +\beta )^2},
\end{align}
respectively. 
The eigenvalue $B_1 = 0$ corresponds to the expected zero mode.
Later, we will confirm that $B_1 = 0$ is indeed the generalized NG mode
and discuss the meaning of the other zero modes for $C_1 = 0$.

One finds a parameter region for which all the nonzero eigenvalues
become positive. For example, if we consider $\beta = \mu =
0,$\footnote{This is also the supersymmetric
limit \cite{Nitta:2017yuf}.} the eigenvalues become 
\begin{align}
&
\mathbf{M}^{00} \ : \ 
A_1 = 0, \quad 
A_2 = \frac{12 \alpha^2 k - 4 \alpha \lambda (\lambda + \sqrt{3 k \alpha
 + \lambda^2})}{9 \alpha^2},
\notag \\
&
\mathbf{M}^{11} \ : \ 
B_1 = 0, \quad 
B_2 = - \frac{4}{3 \alpha} 
\left\{
3 \alpha k + \lambda^2 + \lambda \sqrt{3 \alpha k + \lambda^2}
\right\},
\notag \\
& 
\mathbf{M}^{22} = \mathbf{M}^{33} \ : \ 
C_1 = 0, \quad 
C_2 = \frac{12 \alpha^2 k - 4 \alpha \lambda (\lambda + \sqrt{3 k \alpha
 + \lambda^2})}{9 \alpha^2}.
\label{eq:SUSY_limit_eigenvalues}
\end{align}
One can easily show that all the nonzero eigenvalues $A_2, B_2, C_2$
are positive in  
the region $\alpha < 0$, $\lambda > 0$ and $\lambda^2 + 3 \alpha k > 0$,
and the generalized mass matrix $\mathcal{M}|_0$ is positive semi-definite.
Thus in this region the spatially modulated
configuration \eqref{eq:ansatz_sp} is a 
stable vacuum in the model \eqref{eq:6th_bosonic}.

\paragraph{Generalized NG modes}
Next, we examine the generalized NG modes in the spatially modulated
vacuum \eqref{eq:ansatz_sp}.
The field that has a nonzero VEV is
$\varphi_1 = \del_1 \varphi$.
The VEV is characterized by
\begin{align}
\langle 0 | \varphi_1 | 0 \rangle = v_1 = i c \varphi_0 e^{icx^1}.
\end{align}
As discussed in section \ref{subsect:classification_mod_vac}, the
relevant symmetry breaking of the spatially modulated vacuum is 
$U(1) \times \mathcal{P}^1 \to [U(1) \times \mathcal{P}^1]_{\text{diag}}$. 
The corresponding broken and unbroken generators are given in
Ref.~\cite{Nitta:2017mgk} by 
\begin{align}
T_{\text{b}} = P^1 + T_{U(1)}, \quad 
T_{\text{ub}} = P^1 - T_{U(1)},
\end{align}
respectively, 
where each generator acts on the VEV $v = i c \varphi_0 e^{ic x^1}$ as 
\begin{align}
&
P^1 v_1 = i c \varphi_0 c e^{i c x^1}, \quad 
P^1 \bar{v}_1 = - i c \varphi_0 c e^{- i c x^1}, 
\notag \\
&
T_{U(1)} v_1 = i c \varphi_0 c e^{i c x^1}, \quad 
T_{U(1)} \bar{v}_1 = - i c \varphi_0 c e^{- i c x^1}.
\end{align}
Then, the unbroken generator acts on the VEV as 
$T_{\text{ub}} \vec{v} = 0$.
For the broken generator, we have
\begin{align}
T_{\text{b}} \vec{v} = 2 i c^2 \varphi_0
\left(
\begin{array}{c}
e^{i c x^1} \\
e^{- i c x^1}
\end{array}
\right).
\end{align}

On the other hand, the normalized eigenvector of $\mathbf{M}^{11}|_0$
associated with $B_1 = 0$ is found to be
\begin{align}
&
B_1 = 0 \ : \ 
\vec{u}_1 = \frac{1}{\sqrt{2}} 
\left(
\begin{array}{c}
e^{icx^1}  \\
e^{-icx^1}
\end{array}
\right).
\end{align}
One then finds that
\begin{align}
\vec{u}_1 \propto T_{\text{b}} \vec{v}
\end{align}
is the generalized NG mode.

A comment on the extra zero modes $C_1 = 0$ appearing in the $i=2,3$
sectors is in store.
We have chosen the $x^1$-direction as a specific direction of the spatial
modulation. 
We have then found that the configuration
$\langle \del_1 \varphi \rangle \not= 0$ 
is a (local) minimum of the energy density and there is a
flat direction corresponding to the zero modes of $\mathbf{M}^{11}|_0$.
However, since the model that we consider is Lorentz invariant, the spatial directions
$x^1,x^2,x^3$ are equivalent due to the $SO(3) \in SO(1,3)$ rotational
symmetry. 
Therefore there should be flat directions even in the $\del_2 \varphi,
\del_3 \varphi$-directions in the energy density.
In this sense, the zero modes corresponding to $C_1 = 0$ are
accompanying NG modes to the legitimate generalized NG mode in $x^1$-direction.
On the other hand, the zero mode $A_1 = 0$ in eq.~\eqref{eq:SUSY_limit_eigenvalues} is genuinely accidental since it is
nonzero for general parameters and appears only for the specific choice of
parameters $\mu = \beta = 0$.

\paragraph{Lagrangian for the fluctuation modes}
We will now write down the Lagrangian for the fluctuation modes in the
spatially modulated vacuum.
The Lagrangian \eqref{eq:fluctuation_Lagrangian} becomes
\begin{align}
\mathcal{L} = \frac{1}{2} \sum_{m=0}^3 \vec{\phi}_m^{\dagger}
 \mathbf{L}^{mm}|_0 \vec{\phi}_{m},
\end{align}
where we have ignored the irrelevant constant.
The eigenvalues of $\mathbf{L}^{00}$ are positive and those of
$\mathbf{L}^{ii} = - \mathbf{M}^{ii} \ (i=1,2,3)$ are negative or zero 
in the appropriate parameter region. 
Therefore, the fluctuation fields have the correct signs of their
canonical kinetic terms and  
we expect that there are no dynamical ghost modes in the Lagrangian.
Since $\mathbf{M}^{00} = \mathbf{L}^{00}$, $\mathbf{M}^{ii} = -
\mathbf{L}^{ii}$ (no summation over $i$), the zero modes of $\mathbf{L}^{ij}$ and
$\mathbf{M}^{ij}$ coincide.
We therefore conclude that the kinetic term of the generalized NG mode
vanishes for the spatially modulated vacuum.

As mentioned above, the results in this subsection are essentially the
same as those in Ref.~\cite{Nitta:2017mgk}. 
However, the situation drastically changes when we consider modulations in the
temporal and the light-like directions, which we shall turn to next.

\subsection{Temporal modulation}
For the temporal modulation, we assume the Ansatz for the VEV:
\begin{align}
\langle \del_1 \varphi \rangle = \langle \del_2 \varphi \rangle =
\langle \del_3 \varphi \rangle = 0, \qquad 
\langle \varphi \rangle = \varphi_0 e^{i \omega x^0}.
\label{eq:temp_ansatz}
\end{align}
with real constants $\varphi_0, \omega$.
Assuming this configuration, the matrices $\mathbf{L}^{mn}$ can be
evaluated as 
\begin{align}
\mathbf{L}^{00} =& \  
\left(
\begin{array}{cc}
k + 4 (\lambda + \mu) |\dot{\varphi}|^2 - 9 (\alpha + \beta)
 |\dot{\varphi}|^4
&
2 \dot{\varphi}^2
\left\{
- 3 (\alpha + \beta) |\dot{\varphi}|^2 + (\lambda + \mu)
\right\}
 \\
2 \dot{\bar{\varphi}}^2
\left\{
- 3 (\alpha + \beta) |\dot{\varphi}|^2 + (\lambda + \mu)
\right\}
&
k + 4 (\lambda + \mu) |\dot{\varphi}|^2 - 9 (\alpha + \beta) |\dot{\varphi}|^4
\end{array}
\right),
\notag \\
\mathbf{L}^{11} =& \ \mathbf{L}^{22} = \mathbf{L}^{33} =
\left(
\begin{array}{cc}
- k - 2 \mu |\dot{\varphi}|^2 + (\alpha + 3 \beta) |\dot{\varphi}|^4 
&
- 2 (\dot{\varphi})^2 
\left(
\lambda - \alpha |\dot{\varphi}|^2
\right)
 \\
- 2 (\dot{\bar{\varphi}})^2 
\left(
\lambda - \alpha |\dot{\varphi}|^2
\right)
&
- k - 2 \mu |\dot{\varphi}|^2 + (\alpha + 3 \beta) |\dot{\varphi}|^4 
\end{array}
\right),
\label{eq:temp_L_matrices}
\end{align}
and the others are zero. 
We also have the matrices 
\begin{align}
\mathbf{M}^{00} =& \ 
\frac{\del \mathbf{L}^{00}}{\del \dot{\varphi}} \dot{\varphi}
+
\frac{\del \mathbf{L}^{00}}{\del \dot{\bar{\varphi}}}
 \dot{\bar{\varphi}} + \mathbf{L}^{00}
\notag \\
=& \ 
\left(
\begin{array}{cc}
k + 12 (\lambda + \mu) |\dot{\varphi}|^2 - 45 (\alpha + \beta)
 |\dot{\varphi}|^4 
&
6 \dot{\varphi}^2 
\left[
- 5 (\alpha + \beta) |\dot{\varphi}|^2 + (\lambda + \mu)
\right]
 \\
6 \dot{\bar{\varphi}}^2 
\left[
- 5 (\alpha + \beta) |\dot{\varphi}|^2 + (\lambda + \mu)
\right]
&
k + 12 (\lambda + \mu) |\dot{\varphi}|^2 - 45 (\alpha + \beta) |\dot{\varphi}|^4
\end{array}
\right),
\label{eq:temp_mass1}
 \\
\mathbf{M}^{11} =& \ \mathbf{M}^{22} = \mathbf{M}^{33} = 
\frac{\del \mathbf{L}^{11}}{\del \dot{\varphi}} \dot{\varphi}
+
\frac{\del \mathbf{L}^{11}}{\del \dot{\bar{\varphi}}}
 \dot{\bar{\varphi}} - \mathbf{L}^{11}
\notag \\
=& \ 
\left(
\begin{array}{cc}
3 (\alpha + 3 \beta) |\dot{\varphi}|^4 - 2 \mu |\dot{\varphi}|^2 + k
&
2 \dot{\varphi}^2 
\left(
- \lambda + 3 \alpha |\dot{\varphi}|^2
\right)
 \\
2 \dot{\bar{\varphi}}^2 
\left(
- \lambda + 3 \alpha |\dot{\varphi}|^2
\right)
&
3 (\alpha + 3 \beta) |\dot{\varphi}|^4 - 2 \mu |\dot{\varphi}|^2 + k
\end{array}
\right).
\label{eq:temp_mass2}
\end{align}

Let us analyze the conditions \eqref{eq:eom}, \eqref{eq:energy_extremum}.
Since $\frac{\del \mathcal{L}}{\del (\del_i \varphi)} = 0$ is trivially
satisfied, these conditions reduce to
\begin{align}
\mathbf{L}^{00} 
\left(
\begin{array}{c}
\ddot{\varphi}
 \\
\ddot{\bar{\varphi}}
\end{array}
\right) = 0, \qquad
\mathbf{L}^{00} 
\left(
\begin{array}{c}
\dot{\varphi}
 \\
\dot{\bar{\varphi}}
\end{array}
\right) = 0.
\end{align}
Since $\ddot{\varphi} = i \omega \dot{\varphi}$, $\ddot{\bar{\varphi}} =
- i \omega \dot{\bar{\varphi}}$, the above conditions can be rewritten as 
\begin{align}
\mathbf{L}^{00} 
\left(
\begin{array}{c}
\dot{\varphi}
 \\
- \dot{\bar{\varphi}}
\end{array}
\right) = 0, \qquad
\mathbf{L}^{00} 
\left(
\begin{array}{c}
\dot{\varphi}
 \\
\dot{\bar{\varphi}}
\end{array}
\right) = 0.
\end{align}
Hence, they are satisfied when $\mathbf{L}^{00} = 0$.
This is consistent with condition \eqref{eq:conditions_temp_mod} in
the general discussion.
The condition $\mathbf{L}^{00} = 0$ imposes that the two independent components of $\mathbf{L}^{00}$ in
eq.~\eqref{eq:temp_L_matrices} vanish.
This is possible when the parameter $k$ is chosen as 
\begin{align}
k = - \frac{(\lambda + \mu)^2}{3 (\alpha + \beta)}.
\label{eq:temp_mod_k}
\end{align}
For this choice of parameters, the solution to the condition
$\mathbf{L}^{00} = 0$ is 
\begin{align}
|\dot{\varphi}|^2 = |\omega \varphi_0|^2 = 
\frac{\lambda + \mu}{3 (\alpha + \beta)}.
\label{eq:tm_mod_vac}
\end{align}
Note that since $k$ is the coefficient of the canonical (quadratic)
kinetic term, it should be positive $k>0$. Then we find $ \alpha + \beta< 0$ and $\lambda + \mu < 0$ are necessary for the temporal modulation.

\begin{figure}[t]
\begin{center}
\includegraphics[scale=.6]{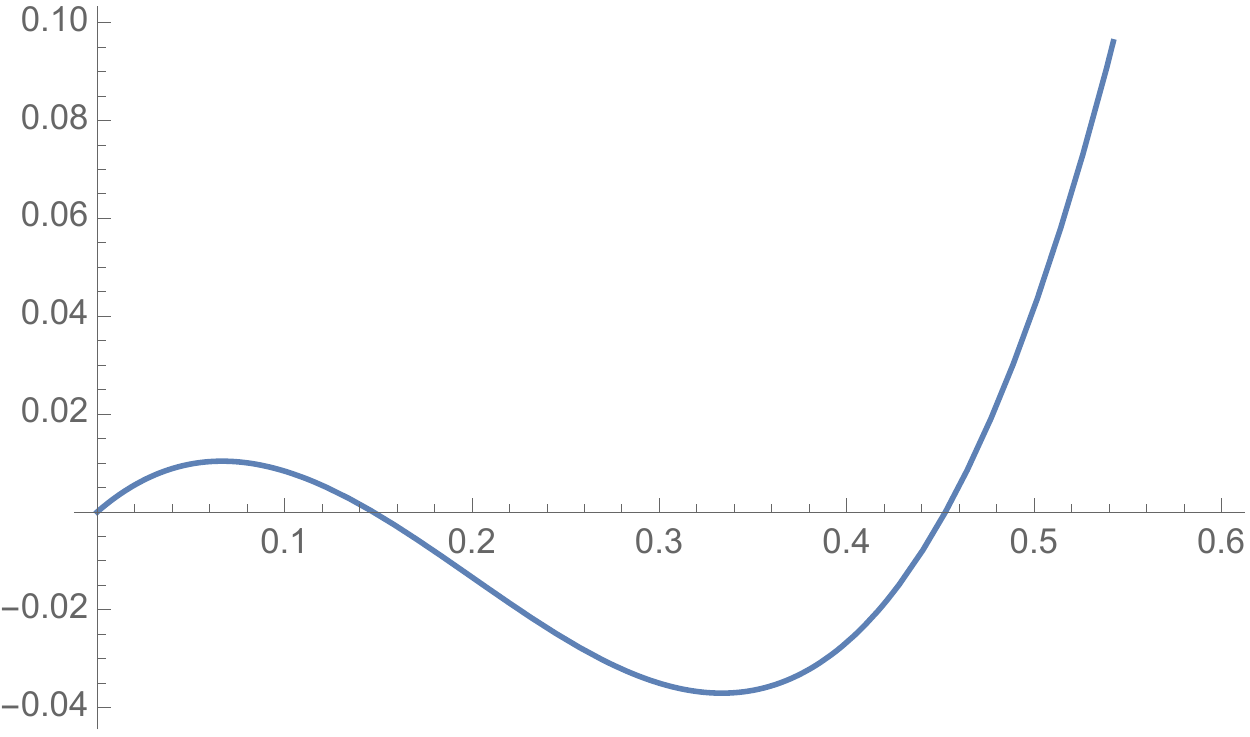}
\end{center}
\caption{The energy density as a function of
 $X=|\dot{\varphi}|^2$. The $X \not= 0$ vacuum is the global minimum.
Here $\alpha + \beta = -1$, $\lambda + \mu = -1$.}
\label{fig:temp_pot}
\end{figure}

\paragraph{Global aspects of the potential and stability of the vacuum}
The energy density for the Ansatz \eqref{eq:temp_ansatz} is
\begin{align}
\mathcal{E}_{\text{temp}} = 
k |\dot{\varphi}|^2 
+ 3 (\mu + \lambda) |\dot{\varphi}|^4 - 5 (\alpha + \beta) |\dot{\varphi}|^6.
\end{align}
This is a function of $X \equiv |\dot{\varphi}|^2 \ge 0$: $y = X \left( k + 3 (\mu +
\lambda) X - 5 (\alpha + \beta) X^2\right)$.
Besides the trivial vacuum $X=0$, there are nontrivial extrema.
The discriminant $\mathbf{D}$ for $y' = k + 6 (\lambda + \mu) X -
15 (\alpha + \beta) X^2 = 0$ is
\begin{align}
\mathbf{D} = 36 (\mu + \lambda)^2 + 90 (\alpha + \beta) k = 16 (\mu +
 \lambda)^2.
\end{align}
This is always positive $\mathbf{D} > 0$. Therefore, there is a local
minimum in the ($X = |\dot{\varphi}|^2$)-direction, as expected. The
solutions to $y' = 0$ are 
\begin{align}
X = |\dot{\varphi}|^2 = \frac{\mu + \lambda}{15 (\alpha + \beta)}, \quad
 \frac{\mu + \lambda}{3 (\alpha + \beta)}.
\end{align}
The former is a local maximum while 
the latter is the temporally modulated vacuum \eqref{eq:tm_mod_vac}.
In order to clarify the (meta)stability of the temporally modulated
vacuum, we examine the zeros of $f$ which is defined by $y = X f$.
Since the discriminant of $f \equiv - 5 (\alpha + \beta) X^2 + 3 (\mu +
\lambda) X + k = 0$ is always positive, 
\begin{align}
\mathbf{D}_f = 9 (\mu + \lambda)^2 + 20 k (\alpha + \beta) = \frac{7}{3}
 (\lambda + \mu)^2 > 0,
\end{align}
the vacuum \eqref{eq:tm_mod_vac} is energetically favored compared with the
trivial vacuum $X = 0$.
Thus, it is the global minimum at least in the
$|\dot{\varphi}|$-direction (see Fig.~\ref{fig:temp_pot}).
We note that the situation is different from the spatial case.
In order that the temporally modulated vacuum is allowed, the
coefficient $k$ of the canonical kinetic term 
should be related to those of the higher-derivative terms via the
relation \eqref{eq:temp_mod_k} and $\alpha + \beta <0$,
$\lambda + \mu < 0$. Therefore, the model is severely constrained and
there are no possibilities for having metastable or degenerate vacua
as in the spatial modulation case.

Since $\frac{\del \mathbf{L}^{0i}}{\del \dot{\varphi}} = 0$, $\frac{\del
\mathbf{L}^{ij}}{\del \dot{\varphi}} = 0 \ (i \not= j)$ for the Ansatz
\eqref{eq:temp_ansatz}, 
we have the mass matrices \eqref{eq:temp_mass1}, \eqref{eq:temp_mass2}.
Hence, the generalized mass matrix is
\begin{align}
\mathcal{M}|_0 = 
\left(
\begin{array}{cccc}
\mathbf{M}^{00}|_0 & & &  \\
& \mathbf{M}^{11}|_0 & & \\
& & \mathbf{M}^{22}|_0 & \\
& & & \mathbf{M}^{33}|_0
\end{array}
\right),
\end{align}
where $k$ is given in eq.~\eqref{eq:temp_mod_k}.
The eigenvalues of $\mathbf{M}^{00}$ are 
\begin{align}
A_1 = 0, \quad A_2 = - \frac{8}{3} \frac{(\lambda + \mu)^2}{\alpha + \beta}.
\end{align}
There is a zero mode and $A_2 > 0$ as expected.
The eigenvalues of $\mathbf{M}^{11}, \mathbf{M}^{22}, \mathbf{M}^{33}$
are 
\begin{align}
B_1 = 0, \quad B_2 = \frac{4 (\lambda +\mu ) (\beta  \lambda -\alpha  \mu )}{3 (\alpha +\beta )^2}.
\end{align}
One can always choose the parameters so that $B_2 > 0$ by choosing
$\alpha + \beta < 0$, $\lambda + \mu < 0$.\footnote{Note that in the
supersymmetric limit $\mu, \beta \to 0$, we 
have an extra zero mode $B_2 = 0$, see Ref.~\cite{Nitta:2017yuf}.}

\paragraph{Generalized NG modes}
The symmetry breaking pattern is essentially the same with the spatial case:
$U(1) \times \mathcal{P}^0 \to [U(1) \times \mathcal{P}^0]_{\text{diag}}$. 
The corresponding broken and unbroken generators are given by
\begin{align}
T_{\text{b}} = P^0 + T_{U(1)}, \quad 
T_{\text{ub}} = P^0 - T_{U(1)},
\end{align}
respectively, 
where each generator acts on the VEV $v_0 = \langle \dot{\varphi} \rangle
= i \omega \varphi_0 e^{i \omega
x^0}$ as
\begin{align}
&
P^0 v_0 = i \omega \varphi_0 \omega e^{i \omega x^0}, \quad 
P^0 \bar{v}_0 = - i \omega \varphi_0 \omega e^{- i \omega x^0}, 
\notag \\
&
T_{U(1)} v_0 = i \omega \varphi_0 \omega e^{i \omega x^0}, \quad 
T_{U(1)} \bar{v}_0 = - i \omega \varphi_0 \omega e^{- i \omega x^0}.
\end{align}
Then, we have $T_{\text{ub}} \vec{v} = 0$, and
\begin{align}
T_{\text{b}} \vec{v} = 2 i c^2 \varphi_0
\left(
\begin{array}{c}
e^{i \omega x^1} \\
e^{- i \omega x^1}
\end{array}
\right).
\end{align}
On the other hand, the normalized eigenvector of $\mathbf{M}^{00}$ 
associated with the zero eigenvalue $A_1 = 0$ is found to be 
\begin{align}
&
 A_1 = 0 \ : \ \vec{u}_1 = \frac{1}{\sqrt{2}} 
\left(
\begin{array}{c}
e^{i \omega x^0}  \\
e^{-i \omega x^0}
\end{array}
\right).
\end{align}
One therefore finds that 
\begin{align}
\vec{u}_1 \propto T_{\text{b}} \vec{v},
\end{align}
is a generalized NG mode.

The zero modes corresponding to $B_1 = 0$ are accidental since 
the normalized eigenvectors of $\mathbf{M}^{11} = \mathbf{M}^{22} =
\mathbf{M}^{33}$ are 
\begin{align}
&
B_1 = 0 \ : \ \vec{u}_1 = \frac{1}{\sqrt{2}} 
\left(
\begin{array}{c}
e^{i \omega x^0}  \\
- e^{-i \omega x^0}
\end{array}
\right).
\end{align}
Hence, the zero modes in
$\mathbf{M}^{11}|_0 = \mathbf{M}^{22}|_0 = \mathbf{M}^{33}|_0$
do not correspond to eigenvectors proportional to the vector generated
by the broken generator $T_{\text{b}} \vec{v}$. 

\paragraph{Lagrangian for fluctuation modes}
For the temporal modulation, there are no direct correspondences between
the fluctuation modes and the zero modes of $\mathcal{M}|_0$, {\it i.e.}
the generalized NG modes.
This is because the matrices $\mathbf{L}^{mn}$ are different from the components
of the generalized matrix $\mathbf{M}^{mn}$.

Since the matrix $\mathbf{L}^{00}$ vanishes in the temporally modulated
vacuum, there is no canonical kinetic term for the fluctuation in the
temporal direction. Only the gradient kinetic term survives in the fluctuation
dynamics.
The Lagrangian \eqref{eq:fluctuation_Lagrangian} becomes,
\begin{align}
\mathcal{L} = \frac{1}{2} \sum_{i=1}^3 \vec{\phi}^{\dagger}_i
 \mathbf{L}^{ii}|_0 \vec{\phi}_i.
\end{align}
In order to clarify whether physically unacceptable modes appear or not, 
we need the eigenvalues of the matrices
\begin{align}
\mathbf{L}^{11}|_0 = \mathbf{L}^{22}|_0 = \mathbf{L}^{33}|_0 = 
\left(
\begin{array}{cc}
- k - 2 \mu |\dot{\varphi}|^2 + (\alpha + 3 \beta) |\dot{\varphi}|^4 & 
- 2 \dot{\varphi}^2 (\lambda - \alpha |\dot{\varphi}|^2)  \\
- 2 \dot{\bar{\varphi}}^2 (\lambda - \alpha |\dot{\varphi}|^2) & 
- k - 2 \mu |\dot{\varphi}|^2 + (\alpha + 3 \beta) |\dot{\varphi}|^4
\end{array}
\right),
\end{align}
where $k = - \frac{(\lambda + \mu)^2}{3 (\alpha + \beta)}$, $|\dot{\varphi}|^2 = \frac{\lambda + \mu}{3 (\alpha + \beta)}$.
The eigenvalues/vectors are found to be 
\begin{align}
&
s_1 = 0, \ : \vec{u}_1 = \frac{1}{\sqrt{2}} 
\left(
\begin{array}{c}
e^{ i \omega x^0} \\
- e^{- i \omega x^0}
\end{array}
\right), 
\notag \\
&
s_2 = \frac{4 (\lambda +\mu ) (2 \alpha  \lambda -\alpha  \mu +3 \beta
 \lambda )}{9 (\alpha +\beta )^2}, 
\ : \ \vec{u}_2 = \frac{1}{\sqrt{2}} 
\left(
\begin{array}{c}
e^{ i \omega x^0} \\
e^{- i \omega x^0}
\end{array}
\right).
\end{align}
We can always choose the parameters so that $s_2 < 0$ by obeying the
conditions $\alpha + \beta < 0$, $\lambda + \mu < 0$.\footnote{For
example, in the supersymmetric limit $\mu = \beta = 0$, we find $s_2
= \frac{8 \lambda^2}{9 \alpha} < 0$ for $\alpha < 0$, which
corresponds to the correct sign of the gradient kinetic
term.}
Therefore we can exclude unphysical ghost modes in appropriate
regions of parameters. 
An alternative discussion of the fluctuations and NG modes in a
time-dependent vacuum can be found in Ref.~\cite{Nicolis:2011pv}.

\subsection{Light-like modulation}
For the light-like modulation, we assume the VEV:
\begin{align}
\langle \del_2 \varphi \rangle = \langle \del_3 \varphi \rangle= 0, \quad 
\langle \del_0 \varphi \rangle = \pm \langle \del_1 \varphi \rangle
 \not= 0,
\qquad 
\langle \varphi \rangle = \varphi_0 e^{i \omega (x^0 \pm x^1)}.
\label{eq:LL_ansatz}
\end{align}
This implies 
\begin{align}
\del \varphi \cdot \del \varphi|_0 = \del \varphi \cdot \del \bar{\varphi}|_0
 = 0.
\end{align}
In the following, we can choose the $+$ sign 
in eq.~(\ref{eq:sp_mod_sol}) without loss of
generality.
With the Ansatz \eqref{eq:LL_ansatz}, the matrices $\mathbf{L}^{mn}$ can be evaluated as 
\begin{align}
\mathbf{L}^{00} =& \ 
\left(
\begin{array}{cc}
k + 4 \lambda |\dot{\varphi}|^2 + 2 \mu |\dot{\varphi}|^2 & 2 \mu
 \dot{\varphi}^2
 \\
2 \mu \dot{\bar{\varphi}}^2 & k + 4 \lambda |\dot{\varphi}|^2 + 2 \mu |\dot{\varphi}|^2
\end{array}
\right), 
\notag \\
\mathbf{L}^{01} =& \  
\left(
\begin{array}{cc}
- 4 \lambda \del_1 \varphi \dot{\bar{\varphi}} - 2 \mu \del_1 \bar{\varphi}
 \dot{\varphi}
&
- 2 \mu \del_1 \varphi \dot{\varphi}
 \\
- 2 \mu \del_1 \bar{\varphi} \dot{\bar{\varphi}} & 
- 4 \lambda \del_1 \bar{\varphi} \dot{\varphi} - 2 \mu \del_1
\varphi \dot{\bar{\varphi}}
\end{array}
\right),
\notag \\
\mathbf{L}^{10} =& \ \mathbf{L}^{\dagger 01} = 
\left(
\begin{array}{cc}
- 4 \lambda \del_1 \bar{\varphi} \dot{\varphi} - 2 \mu \del_1 \varphi
 \dot{\bar{\varphi}}
&
- 2 \mu \del_1 \varphi \dot{\varphi}
 \\
- 2 \mu \del_1 \bar{\varphi} \dot{\bar{\varphi}} & 
- 4 \lambda \del_1 \varphi \dot{\bar{\varphi}} - 2 \mu \del_1
\bar{\varphi} \dot{\varphi}
\end{array}
\right),
\notag \\
\mathbf{L}^{11} =& \ 
\left(
\begin{array}{cc}
- k + (4 \lambda + 2 \mu) |\del_1 \varphi|^2
&
2 \mu (\del_1 \varphi)^2 
 \\
2 \mu (\del_1 \bar{\varphi})^2
&
- k + (4 \lambda + 2 \mu) |\del_1 \varphi|^2
\end{array}
\right), 
\notag \\
\mathbf{L}^{22} =& \ \mathbf{L}^{33} = 
\left(
\begin{array}{cc}
-k & 0 
 \\
0 & -k
\end{array}
\right),
\end{align}
and the others are zero.
We further obtain
\begin{align}
\frac{\del \mathbf{L}^{00}}{\del \dot{\varphi}} =& \ 
\begin{pmatrix}
\dot{\bar{\varphi}} 
\left[
- 8 \alpha |\dot{\varphi}|^2 - 6 \beta |\dot{\varphi}|^2 + 4 (\lambda + \mu)
\right]
&
\dot{\varphi} 
\left[
4 (\lambda + \mu) - 8 \alpha |\dot{\varphi}|^2 - 6 \beta |\dot{\varphi}|^2
\right]
\\
- 6 \beta \dot{\bar{\varphi}}^3
&
\dot{\bar{\varphi}} 
\left[
- 8 \alpha |\dot{\varphi}|^2 - 6 \beta |\dot{\varphi}|^2 + 4 (\lambda + \mu)
\right]
\end{pmatrix},
\notag \\
\frac{\del \mathbf{L}^{01}}{\del \dot{\varphi}} =& \ 
\text{\scriptsize\(
\begin{pmatrix}
4 \alpha |\dot{\varphi}|^2 \del_1 \bar{\varphi}
+ 4 \alpha \dot{\bar{\varphi}}^2 \del_1 \varphi
- 2 \mu \del_1 \bar{\varphi} + 6 \beta |\dot{\varphi}|^2 \del_1 \bar{\varphi}
&
4 \alpha |\dot{\varphi}|^2 \del_1 \varphi 
+ 4 \alpha \dot{\varphi}^2 \del_1 \bar{\varphi}
- 2 \mu \del_1 \varphi + 6 \beta |\dot{\varphi}|^2 \del_1 \varphi
\\
6 \beta \dot{\bar{\varphi}}^2 \del_1 \bar{\varphi}
&
4 \alpha |\dot{\varphi}|^2 \del_1 \bar{\varphi} 
+ 4 \alpha |\dot{\varphi}|^2 \del_1 \bar{\varphi}
- 4 \lambda \del_1 \bar{\varphi} + 6 \beta \dot{\bar{\varphi}}^2 \del_1 \varphi
\end{pmatrix}\)
},
\notag 
\\
\frac{\del \mathbf{L}^{11}}{\del \dot{\varphi}} =& \ 
\text{\scriptsize\(
\begin{pmatrix}
- 4 \alpha \dot{\varphi} (\del_1 \bar{\varphi})^2
- 4 \alpha \dot{\bar{\varphi}} |\del_1 \varphi|^2
- 2 \mu \dot{\bar{\varphi}} - 6 \beta \dot{\bar{\varphi}}^2 |\del_1 \varphi|^2 
&
- 8 \alpha \dot{\varphi} |\del_1 \varphi|^2 
- 4 \lambda \dot{\varphi} - 6 \beta \dot{\bar{\varphi}} (\del_1
 \varphi)^2
\\
- 6 \beta \dot{\bar{\varphi}} (\del_1 \bar{\varphi})^2 
&
- 4 \alpha \dot{\varphi} (\del_1 \bar{\varphi})^2 - 4 \alpha
\dot{\bar{\varphi}} |\del_1 \varphi|^2 - 2 \mu \dot{\bar{\varphi}}
- 6 \beta \dot{\bar{\varphi}} |\del_1 \varphi|^2 
\end{pmatrix}\)
},
\notag \\
\frac{\del \mathbf{L}^{22}}{\del \dot{\varphi}} =& \
\frac{\del \mathbf{L}^{33}}{\del \dot{\varphi}} = 
\begin{pmatrix}
- 2 \mu \dot{\varphi} & 0 
 \\
- 4 \lambda \dot{\bar{\varphi}} & - 2 \mu \dot{\varphi}
\end{pmatrix},
\end{align}
and the others are zero.
Using these results, we can calculate
\begin{align}
\mathbf{M}^{00} &= \ 
\frac{\del \mathbf{L}^{00}}{\del \dot{\varphi}} \dot{\varphi}
+
\frac{\del \mathbf{L}^{00}}{\del \dot{\bar{\varphi}}} \dot{\bar{\varphi}}
+
\mathbf{L}^{00} \non
&=
\begin{pmatrix}
k + (12 \lambda + 10 \mu) |\dot{\varphi}|^2 - (16 \alpha + 12 \beta)
 |\dot{\varphi}|^4 
&
\dot{\varphi}^2 
\left[
4 \lambda + 6 \mu - (8 \alpha + 12 \beta) |\dot{\varphi}|^2 
\right] 
\\
\dot{\bar{\varphi}}^2 
\left[
4 \lambda + 6 \mu - (8 \alpha + 12 \beta) |\dot{\varphi}|^2 
\right] 
&
k + (12 \lambda + 10 \mu) |\dot{\varphi}|^2 - (16 \alpha + 12 \beta)
 |\dot{\varphi}|^4 
\end{pmatrix},
\notag \\
\mathbf{M}^{01} &= \ 
\frac{\del \mathbf{L}^{01}}{\del \dot{\varphi}} \dot{\varphi}
+
\frac{\del \mathbf{L}^{01}}{\del \dot{\bar{\varphi}}} \dot{\bar{\varphi}}
\non
&= 
\begin{pmatrix}
- 2 \mu |\dot{\varphi}|^2 - 4 \lambda |\dot{\varphi}|^2
+ 4 (4 \alpha + 3 \beta) |\dot{\varphi}|^4
&
 \dot{\varphi}^2 
\left[
- 2 \mu + 
(8 \alpha + 12 \beta) |\dot{\varphi}|^2
\right]
 \\
\dot{\bar{\varphi}}^2 
\left[
- 2 \mu + 
(8 \alpha + 12 \beta) |\dot{\varphi}|^2
\right]
&
- 2 \mu |\dot{\varphi}|^2 - 4 \lambda |\dot{\varphi}|^2
+ 4 (4 \alpha + 3 \beta) |\dot{\varphi}|^4
\end{pmatrix},
\notag \\
\mathbf{M}^{11} &= \ 
\frac{\del \mathbf{L}^{11}}{\del \dot{\varphi}} \dot{\varphi}
+
\frac{\del \mathbf{L}^{11}}{\del \dot{\bar{\varphi}}} \dot{\bar{\varphi}}
-
\mathbf{L}^{11} \non
&=
\begin{pmatrix}
k - ( 4 \lambda + 6 \mu) |\dot{\varphi}|^2 - (16 \alpha + 12 \beta)
 |\dot{\varphi}|^4 
&
\dot{\varphi}^2 
\left[
- 4 \lambda - 2 \mu - (8 \alpha + 12 \beta) |\dot{\varphi}|^2 
\right]
 \\
\dot{\bar{\varphi}}^2 
\left[
- 4 \lambda - 2 \mu - (8 \alpha + 12 \beta) |\dot{\varphi}|^2 
\right]
&
k - ( 4 \lambda + 6 \mu) |\dot{\varphi}|^2 - (16 \alpha + 12 \beta)
 |\dot{\varphi}|^4 
\end{pmatrix}, 
\notag 
\end{align}
\begin{align}
\mathbf{M}^{22} &= \ 
\frac{\del \mathbf{L}^{22}}{\del \dot{\varphi}} \dot{\varphi}
+
\frac{\del \mathbf{L}^{22}}{\del \dot{\bar{\varphi}}} \dot{\bar{\varphi}}
-
\mathbf{L}^{22}
=
\left(
\begin{array}{cc}
- 4 \mu |\dot{\varphi}|^2 + k  
& 
- 4 \lambda \dot{\varphi}^2 
\\
- 4 \lambda \dot{\bar{\varphi}}^2 
&
- 4 \mu |\dot{\varphi}|^2 + k  
\end{array}
\right), 
\notag \\
\mathbf{M}^{33} &= \  
\frac{\del \mathbf{L}^{33}}{\del \dot{\varphi}} \dot{\varphi}
+
\frac{\del \mathbf{L}^{33}}{\del \dot{\bar{\varphi}}} \dot{\bar{\varphi}}
-
\mathbf{L}^{33}
=
\left(
\begin{array}{cc}
- 4 \mu |\dot{\varphi}|^2 + k  
& 
- 4 \lambda \dot{\varphi}^2 
\\
- 4 \lambda \dot{\bar{\varphi}}^2 
&
- 4 \mu |\dot{\varphi}|^2 + k  
\end{array}
\right),
\label{eq:LL_mass_matrices}
\end{align}
and the others are zero.

The conditions \eqref{eq:eom}, \eqref{eq:energy_extremum} reduce to 
\begin{align}
&
\mathbf{L}^{00} 
\left(
\begin{array}{c}
\ddot{\varphi}
 \\
\ddot{\bar{\varphi}}
\end{array}
\right)
+ (\mathbf{L}^{01} + \mathbf{L}^{10}) 
\left(
\begin{array}{c}
\del_1 \dot{\varphi}
 \\
\del_1 \dot{\bar{\varphi}}
\end{array}
\right)
+ 
\mathbf{L}^{11} 
\left(
\begin{array}{c}
\del_1^2 \varphi
 \\
\del_1^2 \bar{\varphi}
\end{array}
\right) = 0,
\label{eq:LL1} \\
&
\mathbf{L}^{00} 
\left(
\begin{array}{c}
\dot{\varphi}
 \\
\dot{\bar{\varphi}}
\end{array}
\right) = 0,
\label{eq:LL2} \\
&
\mathbf{L}^{10}
\left(
\begin{array}{c}
\dot{\varphi}
 \\
\dot{\bar{\varphi}}
\end{array}
\right)
- 
\left(
\begin{array}{c}
\frac{\del \mathcal{L}}{\del (\del_1 \bar{\varphi})}
 \\
\frac{\del \mathcal{L}}{\del (\del_1 \varphi)}
\end{array}
\right) = 0.
\label{eq:LL3}
\end{align}
Here we have used $\frac{\del \mathcal{L}}{\del (\del_2 \varphi)} =
\frac{\del \mathcal{L}}{\del (\del_3 \varphi)} = 0$ by assuming the
Ansatz.
Since we have
\begin{align}
\frac{\del \mathcal{L}}{\del (\del_1 \varphi)} = - k \del_1  \bar{\varphi} = 0, \qquad 
\frac{\del \mathcal{L}}{\del (\del_1 \bar{\varphi})} = - k \del_1 \varphi = 0,
\end{align}
the condition \eqref{eq:LL3} becomes
\begin{align}
\left(
\mathbf{L}^{10} + k \mathbf{1}_2
\right)
\left(
\begin{array}{c}
\dot{\varphi}
 \\
\dot{\bar{\varphi}}
\end{array}
\right) = 
\left(
\begin{array}{c}
\dot{\varphi} 
\left(
 k- 4 \lambda |\dot{\varphi}|^2 - 4 \mu |\dot{\varphi}|^2
\right)
 \\
\dot{\bar{\varphi}} 
\left(
 k- 4 \lambda |\dot{\varphi}|^2 - 4 \mu |\dot{\varphi}|^2
\right)
\end{array}
\right) = 0.
\label{eq:LL3-a}
\end{align}
Again, by the Ansatz we have $|\dot{\varphi}|^2 = |\del_1 \varphi|^2$.
Then, the condition \eqref{eq:LL2} becomes
\begin{align}
\left(
\begin{array}{c}
\left(
k + 4 \mu |\dot{\varphi}|^2 + 4 \lambda |\dot{\varphi}|^2
\right) \dot{\varphi}
 \\
\left(
k + 4 \mu |\dot{\varphi}|^2 + 4 \lambda |\dot{\varphi}|^2
\right) \dot{\bar{\varphi}}
\end{array}
\right) = 0.
\label{eq:LL2-a}
\end{align}
These conditions are satisfied when 
\begin{align}
k = \lambda + \mu = 0. 
\label{eq:parameters_LL_conditions}
\end{align}
For the light-like modulation, we have $\dot{\varphi} = \del_1 \varphi$.
Therefore, the condition \eqref{eq:LL1} becomes
\begin{align}
\left(
\mathbf{L}^{00} + \mathbf{L}^{01} + \mathbf{L}^{10} + \mathbf{L}^{11}
\right)
\left(
\begin{array}{c}
\ddot{\varphi}
 \\
\ddot{\bar{\varphi}}
\end{array}
\right) = 0.
\end{align}
However, this condition is automatically satisfied since the matrix
$\mathbf{L}^{00} + \mathbf{L}^{01} + \mathbf{L}^{10} + \mathbf{L}^{11}$
is identically zero by explicit calculations.
Therefore, the light-like modulation is possible when the parameters of
the model satisfy eq.~\eqref{eq:parameters_LL_conditions}.
Evidently, the condition \eqref{eq:parameters_LL_conditions} is weaker
than the condition $k = \lambda = \mu = 0$ determined by the general
discussion in section \ref{subsect:classification_mod_vac}.
The light-like modulation requires $k = \mu + \lambda =
0$ but $\alpha, \beta$ can be nonzero.
Note that for the light-like modulation, the amplitude $\varphi_0$ is
not determined but is a free parameter. 
This is different from the spatial and the temporal modulation cases.

A comment is in order for the light-like modulation.
The model seems to be a little bit strange since the canonical kinetic
term disappears in the Lagrangian \eqref{eq:6th_bosonic}.
Although the Lagrangian contains only the 4th and the 6th order
derivative terms, we will show that the dynamical (fluctuation) field
has quadratic kinetic term in the Lagrangian.

\paragraph{Global aspects of the potential and stability of the vacuum}
For the light-like modulation, the energy density becomes constant:
\begin{align}
\mathcal{E}_{\text{LL}} = 0.
\end{align}
Therefore, there is no ``potential'' when we consider the Ansatz \eqref{eq:LL_ansatz}.
This is why the amplitude $\varphi_0$ is not determined.
The generalized mass matrix is 
\begin{align}
\mathcal{M} = 
\left(
\begin{array}{cc|cc}
\mathbf{M}^{00} & \mathbf{M}^{01} & & \\
\mathbf{M}^{10} & \mathbf{M}^{11} & & \\
\hline
 & & \mathbf{M}^{22} & \\
 & & & \mathbf{M}^{33}    
\end{array}
\right),
\end{align}
where each matrix is obtained from eq.~\eqref{eq:LL_mass_matrices} by
setting $k=\mu+\lambda = 0$ as
\begin{align}
\mathbf{M}^{00} &= \mathbf{M}^{11} = - \mathbf{M}^{01} = -
 \mathbf{M}^{10} \non
 &= 
\begin{pmatrix}
2 \lambda |\dot{\varphi}|^2 - (16 \alpha + 12 \beta) |\dot{\varphi}|^4 & 
\dot{\varphi}^2 
\left[
- 2 \lambda - (8 \alpha + 12 \beta) |\dot{\varphi}|^2
\right]
 \\
\dot{\bar{\varphi}}^2 
\left[
- 2 \lambda - (8 \alpha + 12 \beta) |\dot{\varphi}|^2
\right]
& 
2 \lambda |\dot{\varphi}|^2 - (16 \alpha + 12 \beta) |\dot{\varphi}|^4 
\end{pmatrix},
\notag \\
\mathbf{M}^{22} &= \mathbf{M}^{33} = 
\left(
\begin{array}{cc}
4 \lambda |\dot{\varphi}|^2 & - 4 \lambda \dot{\varphi}^2 \\
- 4 \lambda \dot{\bar{\varphi}}^2 & 4 \lambda |\dot{\varphi}|^2
\end{array}
\right).
\end{align}
In contradistinction to the spatial and the temporal modulations, 
we have nonvanishing off-diagonal blocks
$\mathbf{M}^{01}, \mathbf{M}^{10}$.
The eigenvalues of the submatrix 
\begin{align}
\widetilde{M} \equiv 
\left(
\begin{array}{cc}
\mathbf{M}^{00} & \mathbf{M}^{01}  \\
\mathbf{M}^{10} & \mathbf{M}^{11}
\end{array}
\right),
\end{align}
are 
\begin{align}
A_1 =& \ 0, 
\notag \\
A_2 =& \ 0, 
\notag \\
A_3 =& \ -48 (\omega \varphi_0)^4 (\alpha +\beta ),
\notag \\
A_4 =& \ -8 (\omega \varphi_0)^2 (2 \alpha (\omega \varphi_0)^2 - \lambda ).
\end{align}
The eigenvalues of the matrices $\mathbf{M}^{22} = \mathbf{M}^{33}$ are 
\begin{align}
B_1 = 0, \qquad B_2 = 8 \lambda (\omega \varphi_0)^2.
\end{align}
In order that all the nonzero eigenvalues become positive, we must
require that
\begin{align}
\alpha + \beta < 0, \qquad 
\lambda > 0, \qquad 
2 \alpha \omega^2 \varphi_0^2 < \lambda.
\label{eq:LL_parameters}
\end{align}
It is always possible to choose the parameters such that they satisfy
the above conditions.

\paragraph{Generalized NG modes}
For the light-like modulation, the symmetry breaking pattern is 
$U(1) \times \mathcal{P}^0 \times \mathcal{P}^1 \to [U(1) \times \mathcal{P}^{\pm}]_{\text{diag}}$
where $\mathcal{P}^{\pm}$ represent the translational symmetry group along the light-cone directions
$x^{\pm} = x^0 \pm x^1$. 
The corresponding broken and unbroken generators are given by
\begin{align}
T_{\text{b}} = P^{\pm} + T_{U(1)}, \quad 
T_{\text{ub}} = P^{\pm} - T_{U(1)},
\end{align}
respectively. 
In the following, we choose the $+$ sign in $P_{\pm}$.
The actions of these generators on the VEV $\langle \dot{\varphi} \rangle = \langle \del_1
\varphi \rangle = v_{\text{LL}} = i \omega
\varphi_0 e^{i \omega (x^0 + x^1)}$ are given by
\begin{align}
&
P^{\pm} v_{\text{LL}} = i \omega \varphi_0 \omega e^{i \omega (x^0 +
 x^1)}, \quad 
P^{\pm} \bar{v}_{\text{LL}} = - i \omega \varphi_0 \omega e^{-i \omega (x^0 +
 x^1)},
\notag \\
&
T_{U(1)} v_{\text{LL}} = i \omega \varphi_0 \omega e^{i \omega (x^0 +
 x^1)}, \quad 
T_{U(1)} \bar{v}_{\text{LL}} = - i \omega \varphi_0 \omega e^{-i \omega (x^0 +
 x^1)}.
\end{align}
Thus, we find $T_{\text{ub}} \vec{v} = 0$ and 
\begin{align}
T_{\text{b}} \vec{v} = 2i \omega^2 \varphi_0 
\left(
\begin{array}{c}
e^{i \omega (x^0 + x^1)} \\
e^{- i \omega (x^0 + x^1)} \\
e^{i \omega (x^0 + x^1)} \\
e^{- i \omega (x^0 + x^1)}
\end{array}
\right).
\end{align}

On the other hand, the zero eigenvalues $A_1 = A_2 = 0$ of
$\widetilde{\mathcal{M}}$ are degenerate and whose eigenvectors are 
\begin{align}
& A_1 = 0 \ : \ \vec{u}_1 = 
\left(
\begin{array}{c}
0 \\
1 \\
0 \\
1
\end{array}
\right), \qquad A_2 = 0 \ : \ \vec{u}_2 = 
\left(
\begin{array}{c}
1 \\
0 \\
1 \\
0
\end{array}
\right).
\end{align}
It is clear that the following linear combination of $\vec{u}_1$ and $\vec{u}_2$ defines a zero mode of $\widetilde{\mathcal{M}}$:
\begin{align}
\vec{u}_{12} \equiv \frac{1}{2} 
\left(
e^{- i \omega (x^0 + x^1)} \vec{u}_1 + e^{i \omega (x^0 + x^1)} \vec{u}_2
\right) = 
\left(
\begin{array}{c}
e^{i \omega (x^0 + x^1)} \\
e^{-i \omega (x^0 + x^1)} \\
e^{i \omega (x^0 + x^1)} \\
e^{-i \omega (x^0 + x^1)}
\end{array}
\right).
\end{align}
Hence, we find the generalized NG mode is a zero mode of $\widetilde{M}$:
\begin{align}
\vec{u}_{12} \propto T_{\text{b}} \vec{v}_{\text{LL}}.
\end{align}
One finds that the zero modes corresponding to $B_1 = 0$ are not
proportional to the vector $T_{\text{b}} \vec{v}$ and thus they are
accidental zero modes.

\paragraph{Lagrangian for fluctuation modes}
For the light-like case, since we have $\mathbf{L}^{22}|_0 =
\mathbf{L}^{33}|_0 = 0$, there are no gradient kinetic terms of the
fluctuation in the $m=2,3$ directions.
Instead, we have a cross term in the temporal and the remaining spatial directions.
The Lagrangian \eqref{eq:fluctuation_Lagrangian} becomes 
\begin{align}
\mathcal{L} = \frac{1}{2} \sum_{i,j=0,1} \vec{\phi}_i^{\dagger}
 \mathbf{L}^{ij}|_0 \vec{\phi}_j.
\end{align}
We calculate the eigenvalues of the following matrix:
\begin{align}
\widetilde{\mathbf{L}} \equiv 
\left(
\begin{array}{cc}
\mathbf{L}^{00}|_0 & \mathbf{L}^{01}|_0  \\
\mathbf{L}^{10}|_0 & \mathbf{L}^{11}|_0
\end{array}
\right),
\end{align}
where 
\begin{align}
\mathbf{L}^{00}|_0 = \mathbf{L}^{11}|_0 = 
2 \lambda 
\left(
\begin{array}{cc}
|\dot{\varphi}|^2 & \dot{\varphi}^2 
 \\
\dot{\bar{\varphi}}^2 & |\dot{\varphi}|^2 
\end{array}
\right), \quad 
\mathbf{L}^{01}|_0 = \mathbf{L}^{10}|_0 = - \mathbf{L}^{00}|_0.
\end{align}
The eigenvalues of $\widetilde{\mathbf{L}}$ are 
\begin{align}
s_1 = s_2 = s_3 = 0, \quad s_4 = 8 \lambda (\omega \varphi_0)^2.
\end{align}
Here we have $\omega \varphi_0 = |\dot{\varphi}|$.
Since $\lambda > 0$, we have $s_4 > 0$.
There is only one fluctuation mode whose kinetic term has a positive
coefficient. 
It is, however, not obvious whether the mode is physical or unphysical.
We can say that it is not an unstable mode at least in the parameter
region given by eq.~\eqref{eq:LL_parameters}.

\section{Summary and discussion} \label{sect:conclusion}
In this paper, we have investigated modulated vacua in
Lorentz invariant scalar field theories.  
Although, there are a lot of studies on space- and time-dependent vacua
based on specific models of interest in the literature, we have here
performed a comprehensive analysis on spatial, temporal and light-like
modulated vacua in generic scalar field theories with appropriate
properties.
In particular, our models are in some sense minimal (i.e.~need only
sixth order in derivatives) and are globally \emph{stable}.
We have found the general conditions for modulated vacua.
We also have discussed a generalization of the NG theorem and have
clarified the relation between the zero modes of the generalized mass
matrix and the generalized NG modes.
According to the spontaneous symmetry breaking patterns, the
modulated vacua have been classified into the spatial, temporal and 
light-like ones.
For the spatial and temporal modulations, the diagonal component of the
$U(1)$ and translational symmetries along 
the $x^1$ or $x^0$ direction are preserved.
For the light-like case, the diagonal component of the $U(1)$ and the
translation along the light-cone direction is preserved.
We have presented the general procedure to analyze each modulated
vacuum. 
We have then demonstrated these modulated vacua in a specific model
where up to 6th-order derivative terms of a complex scalar field are
included. 
The model is a generalization of the one considered in
Ref.~\cite{Nitta:2017mgk} where the spatially modulated vacuum was
analyzed. 
We have shown that there are models that allow for temporal and
light-like modulated vacua.
The global structure of
the energy potential in the temporal case is severely restricted,
namely, the temporally modulated state is a global vacuum at least in the
$|\dot{\varphi}|$ direction. 
This is different from the spatially modulated case, for which there
is a parameter region 
where the vacuum is meta-stable, 
degenerate or global minimum.
We have found that the light-like modulation is allowed in the model
where only the quartic and sextic terms remain. We have then 
identified the generalized NG modes in each modulation. 
We have also written down the Lagrangians for the fluctuation modes in
each modulated vacuum. 
As discussed in the previous paper \cite{Nitta:2017mgk}, the kinetic
terms of the generalized NG modes disappear in the spatial
modulation. 
The absence of ghosts is guaranteed due to the positive
semi-definiteness of the generalized mass matrix $\mathcal{M}$.
In the temporal case, the kinetic term of the fluctuation along the
time direction vanishes, but this does not correspond to the
generalized NG mode in general.  
We have found that there is a possible parameter choice for which no
ghosts occur in the temporal modulation.
For the light-like case, the gradient kinetic term along the
$x^2,x^3$-directions are absent while there is only one mode that has
a nonzero kinetic term in the $x^0,x^1$-directions.
There are no obvious conditions for the absence of ghosts.
However, we have pointed out that there is a parameter region where no
ghosts appear in the fluctuation. 

In this paper, we have considered only the FF-type modulation, {\it
i.e.} the modulation of the phase of fields.
It would be interesting to study a possibility of the LO-type where
the amplitude of the fields is modulated. 
This could be realized by including a potential term.
We have studied only the quadratic order of the fluctuations.
Even though, part of the fluctuation modes are absent in each
modulated vacuum, there are possible kinetic terms at higher orders. 
It would be interesting to study the dynamics based on these modes. 

The modulated vacua in a Lorentz invariant setup studied in this paper
are relatively overlooked possibilities of vacuum structures.
As an application of the present discussions,
we will consider a supersymmetric extension of 
the present model, which is possible for $\beta = \mu = 0$, 
and will investigate supersymmetry breaking in modulated
vacua \cite{Nitta:2017yuf,GuNiSaYo}. 
To this end, we can embed our model in higher-derivative chiral models 
formulated in refs.~\cite{Nitta:2014fca,Nitta:2014pwa,Nitta:2015uba}. 
A possibility to use such supersymmetry breaking vacua as a hidden
sector for phenomenological models is an interesting future
direction.

A generalization of the NG theorem in the classical regime has been
discussed in this paper. It would also be important to investigate
the quantum nature of the generalized NG modes.

\subsection*{Acknowledgments}

This work is supported by the Ministry of Education,
Culture, Sports, Science (MEXT)-Supported Program for the Strategic Research Foundation at Private Universities `Topological Science' (Grant No.\ S1511006).
The work of M.~N.~is also supported in part by  
the Japan Society for the Promotion of Science
(JSPS) Grant-in-Aid for Scientific Research (KAKENHI Grant
No.~16H03984 and No.~18H01217).
The work of M.~N.~and S.~B.~G.~is also supported in part by a Grant-in-Aid for 
Scientific Research on Innovative Areas ``Topological Materials
Science'' (KAKENHI Grant No.~15H05855) from the MEXT of Japan. 
The work of S.~S. is supported by JSPS KAKENHI Grant Number JP17K14294.
The work of R.~Y.~is also supported in part by a Grant-in-Aid for
Scientific Research on Innovative Areas 
``Discrete Geometric Analysis for 859 Materials Design'' (KAKENHI Grant No. 17H06462) from the MEXT of Japan.

\end{document}